\documentclass[10pt]{article}

\usepackage[paper=letterpaper,left=1.3in,right=1.3in,top=1in,bottom=1.3in]{geometry}

\usepackage{amsmath, amssymb, amsthm}
\usepackage{xspace}
\usepackage[latin1]{inputenc}
\usepackage[american]{babel}
\usepackage{graphicx}
\usepackage{multirow}
\usepackage[labelfont=bf,font=small,indention=0.3cm,margin=0cm]{caption}  \usepackage[font=footnotesize,margin=5pt,indention=5pt]{subfig}          \usepackage{tikz}
\usepackage{setspace}

\usepackage[numbers,sort&compress,sectionbib]{natbib}             \shortcites{Priezzhev1996}

\newcommand{\RM}{{\ensuremath{\mathcal{R}}}}
\newcommand{\QM}{{\ensuremath{\mathcal{Q}}}}
\newcommand{\LL}{{\ensuremath{\mathcal{L}}}}
\newcommand{\dist}{\operatorname{dist}}
\newcommand{\diam}{\operatorname{diam}}

\newcommand{\poly}{\operatorname{poly}}

\newcommand{\eg}{e.\,g.\xspace}
\newcommand{\uar}{u.\,a.\,r.\xspace}
\newcommand{\kary}{$k$\nobreakdash-ary\xspace}

\newcommand{\x}{{\ensuremath{ \mathbf{x}}}}

\newcommand{\Oh}{\mathcal{O}}

\newcommand{\IGNOREME}[1]{}

\newcommand{\NOTE}[2]{$^{\textcolor{red}\clubsuit}$\marginpar{\setstretch{0.43}\textcolor{blue}{\bf\tiny #1: }\textcolor{red}{\bf\tiny #2}}}
\renewcommand{\NOTE}[2]{}

\newtheorem{thm}{Theorem}  \newtheorem{lem}[thm]{Lemma}

\newtheorem{definition}[thm]{Definition}

\numberwithin{thm}{section}

\newcommand{\thmref}[1]{Theorem~\ref{thm:#1}}

\newcommand{\shortthmref}[1]{Thm.~\ref{thm:#1}}
\newcommand{\shortthmrefs}[2]{Thm.~\ref{thm:#1} and~\ref{thm:#2}}
\newcommand{\lemref}[1]{Lemma~\ref{lem:#1}}
\newcommand{\lemrefs}[2]{Lemmas~\ref{lem:#1} and~\ref{lem:#2}}
\newcommand{\lemrefss}[3]{Lemmas~\ref{lem:#1},~\ref{lem:#2}, and~\ref{lem:#3}}

\newcommand{\defref}[1]{Definition~\ref{def:#1}}

\newcommand{\figref}[1]{Figure~\ref{fig:#1}}

\newcommand{\tabref}[1]{Table~\ref{tab:#1}}

\newcommand{\secref}[1]{Section~\ref{sec:#1}}
\newcommand{\secrefs}[2]{Sections~\ref{sec:#1} and~\ref{sec:#2}}

\newcommand{\eq}[1]{equation~\eqref{eq:#1}}
\newcommand{\eqs}[2]{equations~\eqref{eq:#1} and~\eqref{eq:#2}}

\newcommand{\expanding}{expanding\xspace}

\renewcommand{\Pr}[1]{\ensuremath{\operatorname{\mathbf{Pr}}\left[#1\right]}}
\newcommand{\Pro}[1]{\ensuremath{\operatorname{\mathbf{Pr}}\left[#1\right]}}
\newcommand{\Ex}[1]{\ensuremath{\operatorname{\mathbf{E}}\left[#1\right]}}

\DeclareSymbolFont{AMSb}{U}{msb}{m}{n}
\newcommand{\N}{{\mathbb{N}}}

\newcommand{\R}{{\mathbb{R}}}

\newcommand{\ee}{\varepsilon}

\newcommand{\tsum}{\textstyle\sum}

\let\oldsqrt\sqrt
\def\hksqrt{\mathpalette\DHLhksqrt}
\def\DHLhksqrt#1#2{\setbox0=\hbox{$#1\oldsqrt{#2\,}$}\dimen0=\ht0
   \advance\dimen0-0.2\ht0
   \setbox2=\hbox{\vrule height\ht0 depth -\dimen0}   {\box0\lower0.4pt\box2}}
\renewcommand\sqrt\hksqrt
\renewcommand{\leq}{\leqslant}
\renewcommand{\geq}{\geqslant}
\renewcommand{\le}{\leqslant}
\renewcommand{\ge}{\geqslant}
\renewcommand{\epsilon}{\varepsilon}

\newcount\minute \newcount\hour \newcount\hourMins
\def\now{\minute=\time \hour=\time \divide \hour by 60 \hourMins=\hour \multiply\hourMins by 60
  \advance\minute by -\hourMins \zeroPadTwo{\the\hour}:\zeroPadTwo{\the\minute}}

\def\today{\the\year-\zeroPadTwo{\the\month}-\zeroPadTwo{\the\day}}
\def\zeroPadTwo#1{\ifnum #1<10 0\fi #1}
\title{Quasirandom Rumor Spreading\footnote{This is the final draft (post refereeing) of a paper to appear in the ACM Transactions of Algorithms.
Parts of the results
also appeared in the 19th ACM-SIAM Symposium on Discrete Algorithms (SODA~'08)~\cite{DFS08}
and the 36th International Colloquium on Automata, Languages and Programming (ICALP~'09)~\cite{DFS09}.
\newline
Part of this work was done while Tobias Friedrich and Thomas Sauerwald
were postdoctoral fellows at ICSI Berkeley supported
by the German Academic Exchange Service (DAAD) or
research associates at Max-Planck-Institut f\"ur Informatik.
Benjamin Doerr was partially supported
by project DO 749/4 in the German Research Foundation's (DFG) Priority Program (SPP) 1307.
    \newline
    Authors' addresses:
B.~Doerr,
        Max-Planck-Institut f\"ur Informatik,
        Campus E1 4, 66123 Saarbr\"ucken, Germany;
    T. Friedrich,
   Friedrich-Schiller-Universität Jena,
   Ernst-Abbe-Platz~2,
   07743 Jena, Germany,
        Email: \texttt{friedrich@uni-jena.de};
    T.~Sauerwald,
    Computer Laboratory,
    William Gates Building,
    15 JJ Thomson Avenue,
    Cambridge CB3 0FD, United Kingdom,
    Email: \texttt{thomas.sauerwald@cl.cam.ac.uk}}
}
\author{Benjamin Doerr \and Tobias Friedrich \and Thomas Sauerwald}
\date{}

\begin{document}

\maketitle

\begin{abstract}
    We propose and analyze a quasirandom analogue of the classical push model for
    disseminating information in networks (``randomized rumor spreading'').

    In the classical model, in each round each informed vertex chooses a neighbor
    at random and informs it, if it was not informed before.
    It is known that this simple protocol
    succeeds in spreading a rumor from one vertex to all others
    within $\Oh(\log n)$ rounds on
    complete graphs,
    hypercubes,
    random regular graphs,
    Erd{\H{o}}s-R{\'e}nyi random graph and
    Ramanujan graphs
    with probability $1-o(1)$.
    In the quasirandom model, we assume that each
    vertex has a (cyclic) list of its neighbors. Once informed, it starts at a random
    position on the list, but from then on informs its neighbors in the order of
    the list. Surprisingly, irrespective of the orders of the lists, the above-mentioned bounds still hold.
    In some cases, even better bounds than for the classical model can be shown.

\end{abstract}

\section{Introduction}

\emph{Randomized rumor spreading} or \emph{random phone call protocols} are
simple randomized epidemic algorithms designed to distribute a piece of
information in a network. They build on the basic paradigm that
informed vertices call random neighbors to inform them (\emph{push model}), or
that uninformed vertices call random neighbors to become informed if the neighbor
is (\emph{pull model}). Despite the simple concept, these algorithms succeed in
distributing information extremely fast. In contrast to many natural deterministic
approaches, they are also highly robust against transmission failures \cite{FPRU90,KSSV00,ES06}.

Such algorithms have been applied successfully both in the context where a
single item of news has to be distributed from one processor to all others
(cf.~\cite{Hedetniemi88}), and in the case where news may be injected at various
vertices at different times. The latter problem occurs when maintaining data
integrity in distributed databases, e.g., name servers in large corporate
networks~\cite{DGH+87,KDG03}. For a more extensive, but still concise discussion of
various central aspects of this area, we refer the reader to the paper by \citet{KSSV00}.

\subsection{Randomized Rumor Spreading}

Rumor spreading protocols often assume that all vertices have access to a central
clock. The protocols then proceed in rounds, in each of which each vertex,
independent of the others, can perform certain actions. In the classical
randomized rumor spreading protocols, in each round each vertex contacts a
neighbor chosen independently and uniformly at random. In the push model, which we will focus on here, this results in the contacted vertex becoming informed, provided it was not already. Since all communications are done independently at random, in the following we shall call  this also the {\em fully random model} to distinguish it from the quasirandom one we will propose in this paper.

The first graphs for which the fully random model was analyzed are complete graphs \cite{FG85,Pi87}. \citet{Pi87} proved that with probability $1 - o(1)$, $\log_2 n + \ln n + f(n)$ rounds suffice, where $f(n)$ can be any function tending to infinity.

\citet{FPRU90} showed that on almost all random graphs $\mathcal{G}(n,p)$, $p \ge (1+\ee) \log n/n$, the fully random model runs in $\Oh(\log n)$ time with
probability $1-n^{-1}$. They also showed that this failure probability can be achieved for $p=(\log n + \Oh(\log \log n))/n$ only in $\Omega(\log^2 n)$ rounds. In addition, \citet{FPRU90} also considered hypercubes and proved a runtime bound of $\Oh(\log n)$ with probability $1-n^{-1}$.

For expanders where the maximum and minimum degree satisfy $\Delta/\delta = \Oh(1)$, it was shown in \citet{S10} that the fully random model completes its broadcast campaign in $\Oh(\log n)$ rounds with probability $1-n^{-1}$ (similar results were shown earlier \cite{BGPS06,AS06}, but these hold only for the push-pull model). Recently, \citet{FHP10} and \citet{FP10} derived precise bounds on the runtime for random and pseudo-random regular graphs, extending the result of \citet{FG85} for complete graphs.

\citet{DGH+87} and \citet{KSSV00} introduced the push-pull model which combines push and pull transmissions. For this model, Chierichetti et al.~\citep{CLP10a,CLP10b} and \citet{Giakkoupis11} proved tight runtime bounds in terms of the conductance. In particular, for any graph with constant conductance and arbitrary degree distribution, a runtime bound of $\Oh(\log n)$ was shown in \cite{Giakkoupis11}.

Rumor spreading has recently been studied intensively on social networks,
modeled by random graphs that have a power law degree distribution.
\citet{Chierichetti2009} showed that the push model with non-vanishing probability needs $\Omega(n^\alpha)$ rounds on preferential attachment graphs~\cite{BarabasiAlbert1999} for some $\alpha>0$. For such power-law networks, however, the push-pull
strategy is much better than push or pull alone.
With this strategy, $\Oh(\log n)$ rounds suffice with high probability~\cite{DoerrFF11}.
\citet{DoerrFF11} further proved that for a slightly adjusted process, where contacts are chosen uniformly at random among all neighbors except the one that was chosen just in the round before, $\Oh(\log n/\log\log n)$ rounds suffice.
This is asymptotically optimal as
the diameter of such a preferential attachment graphs, with power law exponent $3$,
is $\Theta(\log n / \log\log n)$~\cite{BRST01}.
\citet{Fountoulakis12} showed that push-pull requires $\Omega(\log n)$
on Chung-Lu-random graphs~\cite{Chung:2002:connected} with power law exponent $>3$ while
for power law exponent $\in(2,3)$, the rumor spreads to almost
all nodes in time $\Theta(\log\log n)$ rounds with high probability.

\subsection{Our Results}

In this work, we propose a quasirandom analogue of the randomized rumor spreading algorithm. In this {\em quasirandom model}, every vertex is equipped with a cyclic list of its neighbors. If a vertex becomes informed, then in the next round it chooses a position on the list uniformly at random and informs the neighbor corresponding to this position. In the subsequent rounds, the vertex continues sending out messages in the order of its list.
Clearly, by introducing these dependencies we gain some natural advantages like the fact that an informed vertex does not call a neighbor a second time before having called all neighbors once. In consequence, we obtain an absolute guarantee that after $\Delta \diam(G)$ rounds all vertices are informed (see Theorem~\ref{thm:degdiambound}) improving over the corresponding $O(\Delta (\diam(G) + \log n))$ bound of \citet{FPRU90} for the fully random model.

Surprisingly, we do not observe that the newly introduced dependencies are harmful. More precisely, we show that the $\Oh(\log n)$ bound (valid with probability $1-n^{-1}$) for complete graphs, hypercubes, random graphs, random regular graphs and Ramanujan graphs in the classical protocol also holds in the quasirandom model \emph{regardless of which lists are used}.
In addition to its theoretical interest, this implies that in an implementation of the quasirandom protocol one may re-use any lists that are already present, e.g., to encode the network structure.

Our $\Oh(\log n)$ runtime bound also applies to very sparse connected random graphs with $p = (\log n+ \omega(1))/n$. This contrasts with a lower bound of $\Omega(\log^2 n)$ steps required by the fully random model to inform all vertices with probability $1-n^{-1}$ \cite[Theorem~4.1]{FPRU90} and with a lower bound on the expected time of $\Omega(\log n \log \log n)$ shown in this paper.
Similarly for hypercubes, we show that the quasirandom model completes in $\Oh(\log n)$ rounds with probability $1-n^{-\Omega(\log n)}$, while the fully random model is easily seen to require $\Omega(\log^2 n)$ steps to achieve the same probability of success. The interesting aspect of these improvements is not so much their actual magnitude, but rather that they can be achieved for free by using a very natural protocol. Note that also speed-ups not visible by asymptotic analyses have been observed, see the experimental analysis~\cite{DoerrFKS11}. For example, the quasirandom protocol was seen to be around 10\% faster on the hypercube on 4096 vertices and around 15\% faster on random $12$-regular graphs on 4096 vertices.

To prove the results in this paper, we need to cope with the more dependent random experiments. Recall that once a vertex has sent out a message, all its future transmissions are determined. The methods we develop to cope with these difficulties, e.g., suitably delaying independent random decisions to have enough independent randomness at certain moments to allow the use of Chernoff-type inequalities, might be useful in the analysis of other dependent settings as well.

Our analysis employs a certain graph class called \emph{expanding graphs},
which is defined by three natural expansion properties.
Roughly speaking, these properties require that small sets of
vertices have many neighbors, and for large sets of vertices the external
vertices have many neighbors in the set, and finally that the vertex degrees are
of similar order (see \defref{ourgraph} for the details).
This graph class has been used by other authors, e.g., in~\cite{CEOR12}.
We prove that complete
graphs, random graphs, random regular graphs and Ramanujan graphs are expanding. After that we show that the quasirandom model succeeds in $\Oh(\log n)$ rounds on every expanding graph with probability $1-n^{-\gamma}$, where $\gamma > 0$ is an arbitrary constant.

\subsection{Related Work on Quasirandomness}

We call an algorithm quasirandom if it imitates (or achieves in an even better way) a particular property
of a randomized algorithm deterministically. The concept of quasirandomness occurs in several areas of
mathematics and computer science. A prominent example are low-discrepancy point sets
and Quasi-Monte Carlo Methods~\cite{niederreiter}, which imitate the property of a random point set to be evenly distributed in their domain.

Our quasirandom rumor spreading protocol imitates two properties of the fully random counterpart, namely that a vertex over a short period of time does not contact neighbors twice and over a long period of time calls all neighbors roughly equally often.

This is very much related to a quasirandom analogue of the classic random walk,
which is also known as
Eulerian walker~\cite{Priezzhev1996},
edge ant walk~\cite{WagnerLB99},
whirling tour~\cite{DumitriuTW03},
Propp machine~\cite{Kleber,CooperSpencer}
and deterministic random walks~\cite{1DPropp,CPC1}.
Unlike in a random walk,
in a quasirandom walk each vertex serves its neighbors in a fixed order.
The resulting (completely deterministic) walk nevertheless closely resembles
a random walk in several respects~\cite{CooperSpencer,CPC1,1DPropp,CooperDFS10,EJC1}.
Other algorithmic applications of the idea of quasirandom walks
are
autonomous agents patrolling a territory~\cite{WLB96},
external mergesort~\cite{BarveGV97},
and iterative load-balancing~\cite{FGS10}.

\subsection{Results Obtained After This Work}

Subsequent to the conference versions~\cite{DFS08,DFS09} and during the
preparation of this journal version, the following results appeared that answer some questions
left open in this work. In~\cite{ADHP09}, it is proven that with probability
$1 - o(1)$, the quasirandom model succeeds in informing all vertices of a complete
graph on $n$ vertices in $(1 + o(1))(\log_2 n + \ln n)$ rounds. Hence for the
complete graph, the quasirandom model achieves the same runtime as the fully
random one~\cite{FG85} up to lower order terms. This was strengthened
by~\citet{FH}, who nearly showed that also Pittel's bounds~\cite{Pi87} hold for the
quasirandom model---their upper and lower bounds deviate by only a $\Theta(\log\log n)$ term.

A second important aspect of broadcasting protocols is their robustness. The
fully random model, due to its high use of independent randomness is usually
considered to be very robust. See~\cite{KSSV00,ES06} for some results in this
direction. A very precise result, valid for both the fully random and the
quasirandom model, was recently given in~\cite{DoerrHL13discmath}. They consider the setting that each message reaches its destination only with an
(independently sampled) probability of $0 < p < 1$. Again for the complete graph on
$n$ vertices, they show that both protocols succeed in $(1 + o(1))\,(\log_{1+p}
n + p^{-1} \ln n)$ rounds with probability $1 - o(1)$. Together with a
corresponding lower bound for the fully random model, this shows that both
models are equally robust against transmission failures, in spite of the greatly
reduced use of independent randomness in the quasirandom model.

The question of how much randomness is needed in such protocols was first considered by \citet{DoerrFouzReducing} and \citet{GW11}. Among other results, the latter work presents a variant of the quasirandom model which requires on average only $\Oh(\log\log n)$ instead of $\Oh(\log n)$ random bits per vertex in order to spread the rumor in $\Oh(\log n)$ rounds on a complete graph with probability $1-n^{-\Omega(1)}$.
\citet{GSSW12} present two protocols that are based on hashing and pseudorandom generators, respectively. While these protocols only require a logarithmic number of random bits in total on many networks, they are more complicated, for instance, they require that random bits are appended to the rumor.

In order to bound the number of messages,~\citet{BES10} analyze another variant of the quasirandom model based on the combination of push and pull calls. This variant is shown to succeed in $\Oh(\log n)$ rounds on random graphs and hypercubes, while requiring only $\Oh(n \,\log \log n)$ messages on random graphs and $\Oh(n\, (\log \log n)^2)$ on hypercubes (all these results hold with probability $1-n^{-1}$).

The worst case behavior of the quasirandom model was very recently addressed by \citet{BFHV12}. Among other results, the authors present a polynomial-time
algorithm to compute the configuration of lists and initial neighbors which maximizes the time to spread the rumor.

\subsection{Organization}

The rest of this paper is organized as follows. In \secref{model} we describe our model more formally and introduce some basic notation. In \secref{general} we derive bounds on the broadcast time that hold for all graphs. After that, in \secref{examined} we describe the class of graphs we consider in this work.
The runtime analysis of quasirandom rumor spreading on this graph class is deferred to \secref{analysis}.
To highlight the efficiency of our new quasirandom model, we also derive some lower bounds for the fully random model in \secref{lowerbounds}. In \secref{hypercube}, we analyze the quasirandom model on hypercubes. We close in \secref{conclusion} with a brief summary of our results.

\renewcommand{\arraystretch}{1.2}
\begin{table*}[bt]
    \footnotesize
    \begin{center}
    \scalebox{0.8}{
    \begin{tabular}{|l|c|c|}
    \hline
    \multirow{2}{*}{\bf Graph class}
    & \multicolumn{2}{c|}{\bf Broadcast time}
    \\
    & \multicolumn{1}{c}{\bf Fully random model}
    & \multicolumn{1}{c|}{\bf Quasirandom model}
    \\
    \hline
    \hline
    \multirow{2}{*}{all graphs}
        & $\Oh(\Delta \,(\diam(G) + \log n))$ \cite{FPRU90}
        & $\leq\Delta\,\diam(G)$ (\shortthmref{degdiambound})
    \\
        & $\leq 12 n \log n$ \cite{FPRU90}
        & $\leq 2n - 3$ (\shortthmref{degdiambound})
    \\
    Complete \kary trees
        & $\Theta(k \log n)$ (\shortthmref{tree})
        & $\Theta(k \log n / \log k)$ (\shortthmref{tree})
    \\
    Hypercubes
        & $\Theta(\log n)$ \cite{FPRU90}
        & $\Theta(\log n)$ (\shortthmref{hypercube})
    \\
    Complete graphs
        & $\Theta(\log n)$ \cite{Pi87,FG85}
        & $\Theta(\log n)$ (\shortthmrefs{randomgraph}{main})
    \\
    Ramanujan
        & $\Theta(\log n)$ \cite{Giakkoupis11}
        & $\Theta(\log n)$ (\shortthmrefs{expander}{main})
    \\
    Almost all random graphs with fixed deg.\ seq.\
        & $\Theta(\log n)$ \cite{Giakkoupis11}
        & $\Theta(\log n)$ (\shortthmrefs{randomreg}{main})
    \\
    Almost all random graphs $G(n,p)$ with
        & \multirow{2}{*}{\vspace{-0.00em} $\Theta(\log^{2} n)$  \cite[Thm.~4.1]{FPRU90}}
        & \multirow{2}{*}{ $\Theta(\log n)$ (\shortthmrefs{randomgraph}{main})}
    \\
    $pn=\log n+\omega(1)$, $pn=\log n+\Oh(\log\log n)$
        &&
    \\
    Almost all random graphs $G(n,p)$ with
        & \multirow{2}{*}{\vspace{-0.00em} $\Theta(\log n)$ \cite{FPRU90}}
        & \multirow{2}{*}{$\Theta(\log n)$ (\shortthmrefs{randomgraph}{main})}
    \\
    $pn=c \log n$, $c > 1$
        &&
    \\
    \hline
    \end{tabular}}
    \end{center}
    \caption{Upper and lower bounds on the broadcast time that hold with probability at least $1-1/n$
             for different graph classes
             in the fully random and the quasirandom model.
             More detailed analyses for
             sparse random graphs
             can be found in
             \tabref{SRG} on page~\pageref{tab:SRG}.
              }
    \label{tab:tab}
\end{table*}

\section{Precise Model and Preliminaries}
\label{sec:model}

Our aim is to spread a rumor in an undirected graph $G=(V,E)$.
Let always $V=\{1,\ldots,n\}$ and $n$ be the number of vertices.
In the quasirandom  model,
each vertex~$v \in V$ is equipped with a cyclic permutation $\pi_v\colon \Gamma(v) \to \Gamma(v)$ of its neighbors $\Gamma(v)$.
We call this its list of neighbors.

The quasirandom rumor spreading process then works as follows. In time step $0$, an arbitrary vertex $s$ is informed initially. If a vertex $v$ becomes informed in time step $t$, then in time step $t+1$ it contacts one of its neighbors $w$ chosen uniformly at random. From then on, it respects the order of the list, that is, in time step $t+1+\tau$, $\tau \in \N$, it contacts vertex $\pi_v^{\tau}(w)$. To simplify the analysis, we will assume that every vertex never stops contacting its neighbors. However, it is easily seen that the propagation of the rumor is exactly the same as if every vertex $v$ stops contacting its neighbors $\deg(v)$ rounds after it got informed. We denote by $I_t$ the set of vertices that are informed at the end of time step $t$.

Note that the assumption that the initial vertex contacted first by an informed vertex is chosen uniformly at random is crucial for the quasirandom protocol. If the adversary was allowed to specify the initial vertices also, then the time to inform all vertices could take up to $n-1$ steps, for example, on a complete graph.

In the remainder of this paper, it will be convenient to consider a model equivalent to the quasirandom model. This model uses the so-called \emph{ever-rolling lists assumption}, where we assume that vertices contact neighbors at all times, informing the neighbors (if the vertex is informed herself). Hence, here each vertex~$v$,
already at the start of the protocol, chooses a neighbor $i_v$
uniformly at random from $\Gamma(v)$. This is the neighbor it contacts at time $t=1$.
In each following time step $t=2,3,\ldots$, the vertex $v$ contacts the vertex $\pi_v^{t-1}(i_v)$ and informs it, if it was not yet informed and if $v$ is informed at that time (here, $\pi_v^{t-1}$ is the $(t-1)$-th composition of $\pi$ with itself).

From the viewpoint of how the information spreads, the model with the ever-rolling lists assumption yields a process equivalent to the standard quasirandom rumor spreading model. Hence in the remainder of the paper, we shall always be discussing the model with ever-rolling lists unless we say otherwise.

We shall analyze how long it takes until a rumor known
to a single vertex is spread to all other vertices.
We adopt a worst-case view in that we aim at bounds that
are independent of the starting vertex and of all lists
present in the model. This suggests the following definitions.
\begin{definition}\label{def:broadcasttime}
    Let $G=(V,E)$ be a graph and $s \in V$. Then by $R_s$ we denote the random variable describing the first time $t$ at which the random rumor spreading process started in the vertex $s$ leads to all vertices being informed. Let $\RM(G)$ be the (unique) minimal integer-valued random variable that dominates all $R_s$, i.e.,
    for every $s \in V$ and $t \in \N$ it holds that
    \[
       \Pro{ \RM(G) \geq t} \geq \Pro{ R_s \geq t}.
    \]
     We call $\RM(G)$ the broadcast time of the randomized rumor spreading protocol on the graph $G$\footnote{In order to see that $\RM(G)$ is well-defined, note that for every $t$ there exists one vertex $s=s(t)$ such that $\Pro{ R_s(G) \geq t}$ is maximized. Then we let $\RM(G)$ satisfy $\Pro{ \RM(G) \geq t} = \Pro{ R_s(G) \geq t}$. Doing this for all integers $t \in \mathbb{N}$ yields a sequence $\{ \Pro{ \RM(G) \geq t} \colon t \in \mathbb{N} \}$ of non-increasing values in $[0,1]$. Hence, $\Pro{ \RM(G) = t} := \Pro{ \RM(G) \geq t} - \Pro{ \RM(G) \geq t+1} $ completes the definition of $\RM(G)$.}.

    Let $\LL = (\pi_v)_{v \in V}$ be a family of lists. By $Q_{\LL,s}$ we denote the (random) first time that the quasirandom rumor spreading protocol with lists $\LL$ started in $s$ succeeds in informing all vertices. Let $\QM(G)$ be the (unique) minimal integer valued random variable that dominates all $Q_{\LL,s}$, i.e., for every family of lists $\LL$, every $s \in V$ and $t \in \N$ it holds that
    \[
       \Pro{ \QM(G) \geq t} \geq \Pro{ Q_{\LL,s} \geq t}.
    \]
     We call $\QM(G)$ the broadcast time of the quasirandom rumor spreading protocol on the graph $G$.
\end{definition}

In the analysis it will often be
convenient to assume that after receiving the rumor, a vertex does not pass it on for a certain number of time steps (\emph{delaying}). Also, it will be helpful to ignore all messages that certain vertices send out from a certain time onward (\emph{ignoring}). Since we assumed all random decisions done by the vertices before the start of the protocol (ever-rolling list assumption), an easy induction shows that any delaying and ignoring assumptions (possibly even relying on the random choices done by the vertices which have not been active yet) for each vertex can only increase the round in which it becomes informed. In consequence, these assumptions can only increase the time needed to inform all vertices. More precisely, the random variable describing the broadcast time of any model with delaying and ignoring assumptions dominates the original one (see \defref{dominance} for the precise definition of stochastic domination).

\begin{lem}\label{lem:delay}
    For all possible delaying and ignoring assumptions, the random variable describing the broadcast time
    of the quasirandom model with these assumptions is stochastically larger than
    the broadcast time of the true quasirandom model.
\end{lem}

We use both delaying and ignoring to reduce the number of dependencies in the analysis. We do this by splitting the analysis into \emph{phases}. All vertices that receive the rumor within this phase (\emph{newly informed vertices}) are assumed to delay their actions until the beginning of the next phase. From this next phase on, all messages from vertices that previously sent out messages are ignored. Thus, we start each phase with only newly informed vertices acting. Since they have not actively participated in the rumor spreading process, the first neighbors to which they send the rumor are chosen independently.

\label{def:reach}
We will also need chains of contacting vertices.
That is, we say a vertex~$u_1\in V$ \emph{reaches} another vertex~$u_m\in V$ within the
time interval $[a,b]$, if there is a path
$(u_{1}, u_{2}, \dots,  u_{m})$ in $G$ and
$t_{1} < t_{2} < \dots < t_{m-1} \in [a,b]$ such that
for all $j \in [1,m-1]$, $\pi_{u_{j}}^{t_j-1}(i_{u_j}) =
u_{j+1}$. For a vertex $w \in V$, we denote by $U_{[a,b]}(w)$  the set of vertices that reach $w$ within the time interval $[a,b]$.

\subsection*{Other Notation}

Throughout the paper, we use the following graph-theoretical notation.
For a vertex $v$ of a graph $G=(V,E)$, let $\Gamma(v) :=\{u \in V \colon
\{u,v\} \in E\}$ be the set of its \emph{neighbors} and
$\deg(v):= |\Gamma(v)|$ its \emph{degree}.
For any $S \subseteq V$, let $\deg_S(v):=|\Gamma(v)\cap S|$.
For any $S_1,S_2 \subseteq V$, let $E(S_1,S_2):=\{(u,v)\in E \colon u\in S_1 \wedge v\in S_2\}$.
Let
$\delta:=\min_{v\in V} \deg(v)$ be the \emph{minimum degree},
$d:=2|E|/n$ the \emph{average degree},
and
$\Delta:=\max_{v\in V} \deg(v)$ the \emph{maximum degree}.
The \emph{distance} $\dist(x,y)$ between vertices $x$ and $y$ is the length of a shortest path from $x$ to~$y$.
The \emph{diameter} $\diam(G)$ of a connected graph $G$ is
the largest distance between two vertices in $G$.
We will also use
$\Gamma^k(u) := \{v \in V \colon \dist(u,v) = k\}$
and
$\Gamma^{\leq k}(u) := \{v \in V \colon \dist(u,v) \leq k\}$.
For sets $S$ we define $\Gamma(S) := \{v \in V \colon \exists u\in S, \{u,v\} \in E \}$
as the set of neighbors of $S$.
The complement of a set $S$ is
denoted
$S^c:=V \setminus S$.

All logarithms $\log n$ are natural logarithms to the base $e$.
As we are only interested in the asymptotic behavior,
we will sometimes assume that $n$ is sufficiently large.

\section{Quasirandom Rumor Spreading on General Graphs}
\label{sec:general}

In this section, we prove two bounds for the broadcast time valid for all graphs.
The corresponding upper bounds for the fully random model are
$\Oh(\Delta \, (\diam(G) + \log n))$ and
$12 n \log n$, both satisfied with probability $1-1/n$~\cite{FPRU90}.

\begin{thm}\label{thm:degdiambound}
    For any graph $G=(V,E)$, the broadcast time of the quasirandom model is at most
    \begin{enumerate}
        \setlength{\itemsep}{0pt}
        \setlength{\parskip}{0pt}
        \item $\Delta\cdot\diam(G)$ with probability~1, and
        \item $2n - 3$ with probability~1.
    \end{enumerate}
\end{thm}
\begin{proof}
    Let $u$ be the vertex initially informed.

    Let $v \in V$ and $P=(u=u_0,u_1,\ldots,u_\ell=v)$ be a shortest path from $u$ to~$v$.
    Clearly for all $i \le \ell$, $u_i$ becomes informed at most $\deg(u_{i-1}) \leq \Delta$ time-steps after $u_{i-1}$ became informed.
    Claim \emph{(i)} follows.

    To prove claim \emph{(ii)},
     again let $v \in V$ and let $P = (u = u_0, u_1, \ldots, u_\ell = v)$ be a shortest
    path from $u$ to~$v$.
    Let $w$ be a vertex not lying on $P$.
    Then, as observed already in~\cite{FPRU90}, $w$ has at most three neighbors on $P$, and these
    are contained in $\{u_{i-1}, u_i, u_{i+1}\}$ for some $i < \ell$. If
    $w$ has exactly three neighbors $u_{i-1}, u_i,
    u_{i+1}$ on $P$, we call it a counterfeit of~$u_i$ (as $u_i$ and $w$ have,
    apart from each other, the same neighbors on $P$). Denote by
    $C(u_i)$ the set of counterfeits of~$u_i$. Without loss of generality, we
    may choose $P$ in such a way that for all $i < \ell$, $u_i$ is
    informed no later than any if its counterfeits.

    Note also that any vertex~$u_i$ on the path has only $u_{i-1}$ and $u_{i+1}$
    (if existent) as neighbors on the path.

    Let $t_i$ denote the time that vertex~$u_i$ becomes informed. Then,
    $t_0 = 0$. By definition of our algorithm and choice of~$P$, we have
    $t_1 \le t_0 + |\Gamma(u_0) \setminus C(u_1)| = t_0 + |\Gamma(u_0)\setminus P| +
    1 - |C(u_1)|$. For $2 \le i \le \ell-1$, similarly, we have $t_i \le t_{i-1}
    + |\Gamma(u_{i-1}) \setminus C(u_i)| =   t_{i-1} + |\Gamma(u_{i-1})\setminus P| +
    2 - |C(u_i)|$. Finally, $t_\ell \le t_{\ell-1} + |\Gamma(u_{\ell-
    1})\setminus P| + 2$. We conclude
    \[
        t_\ell \le \sum_{i = 0}^{\ell-1} |\Gamma_{V}(u_i)  \setminus P| - \sum_{i = 1}^{\ell-1} |C(u_i)| + 2 \ell - 1.
    \]
    Now each vertex~$w$ not lying on $P$ can contribute at most $2$ to the above
    expression (if it has three neighbors on $P$, then it is also a
    counterfeit). Hence $t_\ell \le 2 (n - \ell -1) + 2 \ell - 1 = 2n - 3$.
\end{proof}

It is easy to verify that for a path of length $n-1$ there are lists and initial vertices such that $2n-3$ rounds are needed. Hence the second bound is tight. The first bound is matched by $k$-ary trees (up to constant factors), as shown in \secref{trees}, where we also demonstrate that the quasirandom model is faster than the fully random one on these graphs.

\section{Graph Classes}\label{sec:examined}

Our results cover hypercubes, many expander graphs, random regular graphs,
and Erd{\H{o}}s-R{\'e}nyi random graphs.  The three latter graph classes
have three properties in common, to which we will refer as ``expanding''.
This allows us to examine the quasirandom rumor spreading on them
from a higher level just using these three properties defined in the following
\secref{expanding}.

\newcommand{\propexpvertex}{\textup{\bf{(P1)}}\xspace}
\newcommand{\propexpedge}{\textup{\bf{(P2)}}\xspace}
\newcommand{\propdeg}{\textup{\bf{(P3)}}\xspace}
\newcommand{\propdegg}{\textup{\bf{(P3')}}\xspace}

\subsection{Expanding Graphs}
\label{sec:expanding}

In order to analyze our quasirandom rumor spreading model for
a larger class of graphs at once,
we distill three simple properties of graphs
which are satisfied by several common graph classes.
Given these three properties, we can later
prove in \thmref{main} that quasirandom rumor spreading
successfully informs all vertices in a logarithmic runtime.
Roughly speaking, these properties concern the vertex expansion of
not too large subsets \propexpvertex, the edge expansion \propexpedge and
the regularity of the graph \propdeg.

\begin{definition}[expanding graphs]
\label{def:ourgraph}
We call a connected graph \emph{\expanding} if the following properties hold:
\begin{description}
    \item[\propexpvertex]
        For any constant $C_\alpha$ with
        $0<C_\alpha\leq d/2$
        there is a constant $C_\beta\in(0,1)$
        such that for any connected subset $S \subseteq V$ with $3 \leq |S| \leq C_{\alpha} \, (n/d)$,
        it holds that
        $ |\Gamma(S) \setminus S| \geq  C_{\beta} \, d \, |S|$.
    \item[\propexpedge]
        There are constants $C_\delta\in (0,1)$ and $C_{\omega} > 0$
        such that for any subset $S \subseteq V$,         the number of vertices in $S^c$ which have at least $C_\delta d(|S|/n)$
        neighbors in $S$ is at least $|S^c| - \frac{C_{\omega} n^2}{d|S|}$.
                      \item[\propdeg]
        $d=\Omega(\Delta)$ and if $d=\omega(\log n)$, then also $d=\Oh(\delta)$.
\end{description}
\end{definition}

We will now describe the properties in detail and
argue why each of them is intrinsic for the analysis.
\propexpvertex\ describes a vertex expansion, which means
that connected sets have a neighborhood which
is roughly in the order of the average degree larger than the set itself.
Without this property, the broadcasting process could end up in
a set with a tiny neighborhood and thereby slow down too much.
Note that in \propexpvertex, $C_\beta$ depends on $C_\alpha$.
As $C_\alpha$ has to be a constant,
the upper limit on $C_\alpha$ only applies for constant $d$.

\propexpedge\ is a certain edge expansion property
implying that a large portion of uninformed vertices has
a sufficiently large number of informed neighbors. This avoids
the situation where the broadcasting process stumbles upon a point when it
has informed many vertices but most of the remaining uninformed
vertices have very few informed neighbors and therefore only
a small chance to get informed. Note that \propexpedge\ is only useful
for $|S|=\omega(n/d)$.

The last property \propdeg\ demands a certain regularity of the
graph.  It is trivially fulfilled for regular graphs,
which many definitions of expanders require.
The condition $d=\Omega(\Delta)$ for the case $d=\Oh(\log n)$
does not limit any of our graph classes below.
If the average degree is at most logarithmic, \propdeg\ implies
no further restrictions.  Otherwise, we require $\delta$, $d$ and $\Delta$
to be of the same order of magnitude.  Without this condition, there could
be an uninformed vertex with $\delta$ informed neighbors of degree $\omega(\delta)$
which does not get informed in logarithmic time with a good probability.
With an additional factor of $\Delta/\delta$ this could be resolved,
but as we aim at a logarithmic bound, we require $\delta=\Theta(\Delta)$
for $d=\omega(\log n)$.
Note that we do not require $d=\omega(1)$, but the proof techniques
for constant and non-constant average degrees will differ in \secref{analysis}.

We now describe several important graph classes which are \expanding,
i.e., satisfy all three properties of \defref{ourgraph},
with high probability.

\subsubsection{Complete Graph}

It is not difficult to show that complete graphs are \expanding.

\begin{thm}
    Complete graphs are \expanding.
\end{thm}
\begin{proof}
    We first prove that \propexpvertex holds. Let $C_{\alpha}$ be an arbitrary
    constant. Take any subset $S \subseteq V$ with $3 \leq |S| \leq C_{\alpha}
    n/(n-1) $. Then
    \[
        |\Gamma(S) \setminus S|
        = n - |S|
        \geq |S| \, (n-1) \, \frac{n - |S|}{|S| \, n}
        = |S| \, (n-1) \, \left( \frac{1}{|S|} - \frac{1}{n} \right),
    \]
    so \propexpvertex holds with $C_{\beta}=\frac{1}{|S|} - \frac{1}{n} \geq \frac{n-1}{C_{\alpha} n} - \frac{1}{n} > 0$.
        We now show that \propexpedge holds. Let $C_{\delta} \in (0,1)$ be an arbitrary
    constant. Take any subset $S \subseteq V$. Then every vertex $v \in S^c$ has
    exactly $|S| \geq C_{\delta} d (|S|/n)$ neighbors in $S$ which implies that
    \propexpedge is satisfied.

    Property \propdeg is trivially fulfilled, as a complete graph is regular.
\end{proof}

\subsubsection{Random Graphs $\mathcal{G}(n, p)$, $p \geq (\log n + \omega(1))/n$}
\label{sec:randomgraph}

In this section we show that a large class of random graphs
is expanding with probability \mbox{$1-o(1)$}.
We use the popular random graph model $\mathcal{G}(n,p)$, where between each two vertices out of a set of $n$ vertices an edge is
present independently with probability~$p$. This model is usually
called the Erd\H os-R\'enyi random graph model.

We distinguish two kinds of random graphs with slightly different properties:
\begin{definition}[sparse and dense random graph]
    We call a random graph $\mathcal{G}(n,p)$ \emph{sparse} if
    $p=(\log n + f_n)/n$ with $f_n=\omega(1)$ and $f_n=\Oh(\log n)$, and
\emph{dense} if $p=\omega(\log(n)/n)$.
\end{definition}
Note that our definition of a sparse random graph coincides with the one
of \citet{CF07} who set $p = c_n \log (n)/n$
with $(c_n-1)\log n=\omega(1)$ and $c_n = \Oh(1)$.
In the remainder of this section we prove the following theorem.

\begin{thm}
\label{thm:randomgraph}
Sparse and dense random graphs are \expanding with probability \mbox{$1-o(1)$}.
\end{thm}

The proof can be skipped at a first reading of the paper, since the following
sections do not depend on the proven results of this section.
\begin{proof}
Note that for random graphs, $d=p\, (n-1)\, (1\pm o(1))$ holds with probability $1-n^{-1}$.
To simplify the presentation of the proof we will ignore
the factor $(1\pm o(1))$ as we do not try to optimize the used constants.

The easiest property to check is \propdeg.
That $d=\Omega(\Delta)$ holds with probability $1-o(1)$ is a well-known property of random graphs and
can be shown by union and Chernoff bounds (cf.~\lemref{chernoff}) as follows:
\[
    \Pr{\Delta \geq 5d}
        = \Pr{\exists v\in V \colon \deg(v)\geq 5d}
        \leq n\exp(-4d/3)
        = o(1).
\]
Analogously for $d=\omega(\log n)$,
\[
    \Pr{\delta \leq d/2}
        = \Pr{\exists v\in V \colon \deg(v)\leq d/2}
        \leq n\exp(-d/8)
        = o(1).
\]

\noindent
For the proof of \propexpedge\ it suffices to bound the number of neighbors
of a set by Chernoff bounds.
The following lemma does this for sparse and dense random graphs at once.

\begin{lem}
    \label{lem:randomgraph:propexpedge}
    Sparse and dense random graphs satisfy \propexpedge\ with probability \mbox{$1-o(1)$}.
\end{lem}
\begin{proof}
    We choose $C_\delta=1/2$ and $C_\omega=32$.
    Consider a set $S \subseteq V$ of arbitrary size $|S|=s$.
    We want to show that the number of
    vertices in $S^c$ which have at least $C_\delta ds/n$ neighbors in $S$ is at
    least $|S^c| - C_\omega \frac{n^2}{ds}$.

    Fix a vertex $v \in S^c$. Linearity of expectations implies
    $
      \Ex{ \deg_{S}(v) } = \sum_{u \in S} p = ps$.
    Hence a Chernoff bound (\lemref{chernoff}) gives
    \begin{align*}
      \Pr{ \deg_{S}(v) \leq (1/2) \Ex{ \deg_{S}(v) } }
      &\leq \exp \left( - \frac{ds}{8n} \right).
    \end{align*}
    Hence the probability
    for the existence of a subset of vertices in $S^c$ of size $C_\omega n^2/(ds)$ being {\em bad}, i.e.,
    the set has more than $\frac{C_\omega n^2}{ds}$ vertices with less than $C_\delta ds/n$ neighbors in
    $S$, can be bounded by
    \begin{align*}
       \binom{n-s}{\frac{C_\omega n^2}{ds}} \, \exp\left( - \frac{ds}{8n} \right)^{C_\omega n^2/(ds)}
      &\leq 2^n \, \exp(-4 n).
    \end{align*}
    Taking the union bound over all possible sets $S$, we obtain
    \[
    \Pr{ \exists \, \text{bad $S$}}
    \leq 2^n \cdot 2^n \, \exp(-4 n)
        \leq \left( \frac{4}{e^4} \right)^n.
    \qedhere
    \]
     \end{proof}

We now turn to \propexpvertex.
We first prove that \propexpvertex\ holds for dense random graphs.
After that we extend it to sparse random graphs,
which requires slightly more involved arguments.
\begin{lem}
    \label{lem:denserandomgraph:propexpvertex}
    Dense random graphs satisfy \propexpvertex\ with probability \mbox{$1-o(1)$}.
\end{lem}
\begin{proof}
    Let $C_\alpha>0$ be an arbitrary constant.
    Fix a set $S\subseteq V$ of size $s=|S|$ with $1 \leq s \leq C_\alpha (n/d)$.
    We show that
    $|\Gamma(S) \setminus S| \geq C_\beta d s$ with
    $C_\beta := 1/(4(C_\alpha+1))$.

    The probability that a vertex
    $v \in S^{c}$ is connected to a vertex in $S$ is
    \[
        1 - \left(1 - p \right)^{s} \geq 1 - \exp(-ps).
    \]
    Linearity of expectation     and using the fact that $e^{-x} \leq \frac{1}{x+1}$ for any number $x\geq0$
    gives
    \begin{align*}
        \Ex{|\Gamma(S) \setminus S|}
                &\geq (n-s) \, \big(1 - \tfrac{1}{ps+1} \big) \\
        &= \big(n-o\big(\tfrac{n}{\log n}\big)\big) \, \tfrac{ps}{ps + 1}
        \geq \tfrac{n}{2} \, \tfrac{ps}{C_\alpha+1} = 2\,C_\beta ds.
    \end{align*}

    \noindent
    Applying Chernoff bounds (\lemref{chernoff}), we obtain
    \begin{align*}
        \Pr{|\Gamma(S) \setminus S| \leq C_\beta\,ds }
        \leq \exp \left( - C_\beta ds/4 \right).
    \end{align*}

    \noindent
    It remains to show that this holds for all sets $S$.
    First,
    taking a union bound over all sets of size $s$, we obtain
    \[
        \Pr{ \exists \, S \subseteq V\colon
               |S|=s,\;|\Gamma(S) \setminus S| \leq C_\beta\,ds }
        \leq n^{s} \,  \exp \left( -C_\beta ds/4 \right)
        \leq n^{-\omega(1)},
    \]
    where the last inequality uses the assumption $d = \omega(\log n)$.
    Finally, a union bound over all possible values of $s$ yields
    \[
        \Pr{ \exists \, S \subseteq V\colon
             |\Gamma(S) \setminus S| \leq C_\beta\,ds }
        \leq \tsum_{s=1}^{n} n^{-\omega(1)} = n^{-\omega(1)}.
        \qedhere
    \]
\end{proof}

\noindent
We now consider sparse random graphs.  For this,
we need the following three technical lemmas.  The first one
proves a slightly stronger bound compared
to the original lemma in
\cite[Property P2]{CF07}.
\begin{lem}\label{lem:sparserandomgraph:propexpvertex:help1}
    Sparse random graphs satisfy
    with probability \mbox{$1-o(1)$} that
    for every subset $S \subseteq V$ of size $s = \Oh(n/d)$ it holds that
    $|E(S,S)| = o(s \log n)$.
\end{lem}
\begin{proof}
    We assume without loss of generality $S\neq\emptyset$.
    We bound the probability for
    the existence of a set $S$ of size $s$
    with $|E(S,S)| \geq s \frac{\log n}{\sqrt{\log \log n}}$ as follows:
    \begin{align*}
    &\qquad\Pr{ \exists S \colon |E(S,S)| \geq s \frac{\log n}{\sqrt{\log \log n}} }\\
    &\qquad\leq \binom{n}{s} \,
          \binom{ \binom{s}{2} }{ s \frac{\log n}{\sqrt{\log \log n}}}\,
          p^{s \frac{\log n}{\sqrt{\log \log n}}} \\
    &\qquad\leq n^s \,
          \left(  \frac{s^2 \, e}{ s \frac{\log n}{\sqrt{\log \log n}}} \right)^{ s \frac{\log n}{\sqrt{\log \log n}}}\!
          p^{s \frac{\log n}{\sqrt{\log \log n}}}
    = n^s \,
           \left( \frac{s\,e\,p\, \sqrt{\log \log n}}{ \log n}  \right)^{ s \frac{\log n}{\sqrt{\log \log n}}}\\
    &\qquad= \exp \left( -  s \left( \frac{\log n}{\sqrt{\log \log n}} \,
        \log \left( \frac{ \log n  }{ s\,e\,p\, \sqrt{\log \log n}}  \right)
        - \log n\right)
        \right)\\
    &\qquad\leq \exp \left( - \Omega(\log n \, \sqrt{\log \log n}) - \log n\right)\\
    &\qquad= n^{-\omega(1)},
    \end{align*}
where in the third inequality we used that $s = \Oh( n / d)$ and $p = \Theta(d/n)$ together imply that $s\,e\,p = \Oh(1)$. Taking the union bound over all values of $s$ completes the proof.
\end{proof}

\noindent
It is known that in very sparse random graphs, vertices
with small degree are rare and far away.  To prove
\propexpvertex\ we need the following statement.
\begin{lem}
    \label{lem:randomgraph:propsmalldist}
    Sparse and dense random graphs satisfy
    with probability \mbox{$1-o(1)$} that
    no two vertices of degree at most $d/50$ are within distance at most 3.
\end{lem}
\begin{proof}
    We will prove a slightly stronger statement, that is, there are no two vertices
    of degree at most $d/50$ within distance at most \mbox{$\log(n) / (\log\log n)^2$} with probability $1-o(1)$.

    For $d\leq 2.5 \log n$ we use
    property P2 of Lemma~1 of \citet{CF07} which states that
    no two vertices of degree at most $\log n/20$ are within distance at most
    $\log(n) / (\log\log n)^2$ with probability $1-o(1)$.

    For $d\geq 2.5\log n$ we calculate by Chernoff bounds
    that the probability that an arbitrary vertex has at most $d/50$ neighbors
    is
    $
        \exp\left( - (49^2\,d)/(2\cdot50^2)\right)
        \leq n^{-1.2}$.
    Therefore the probability that there exists a vertex with at most $d/50$ neighbors
is
    $
        n\cdot n^{-1.2} = o(1)$
            and the claim is satisfied.
\end{proof}

We also need the following simple graph-theoretical lemma. We shall use it later with $d$ being the average degree, but it holds for $d$ being an arbitrary number.
\begin{lem}\label{lem:sparserandomgraph:propexpvertex:help2}
    Let $d \in \N$ and $G$ be a graph where
    no two vertices of degree at most $d/50$ are within distance at most $2$.
    Then for any connected $S \subseteq V$ having at least two vertices,
    $\sum_{v \in S} \deg(v) \ge (d/100)|S|$.
        \end{lem}
\begin{proof}
    Call a vertex \emph{small} if it has degree less than $d/50$, otherwise we call it \emph{big}.
        Let $T$ be a spanning tree of $S$. Let $x$ be any vertex in $S$ that is not small, i.e., big. For any small vertex $u \in S$, let $\pi(u)$ be the unique
    neighbor of $u$ that is on the unique path from $u$ to $x$ in $T$. Since two
    small vertices have distance at least three, $\pi(u)$ is big, and for
    different small vertices $u_1, u_2$, we have $\pi(u_1) \neq \pi(u_2)$. Hence
    $\pi$ is an injective mapping of small vertices into big vertices. In consequence,
    $S$ contains at least $|S|/2$ big vertices. Hence  $\sum_{v \in S} \deg(v)
    \ge (|S|/2)(d/50) = (d/100)|S|$.
\end{proof}

Using all three above lemmas, we prove \propexpvertex\ for sparse graphs.
\begin{lem}\label{lem:sparserandomgraph:propexpvertex}
    Sparse random graphs satisfy \propexpvertex\ with probability \mbox{$1-o(1)$}.
\end{lem}
\begin{proof}
    To prove \propexpvertex, let $C_\alpha>0$ be an arbitrary constant
    and let $S\subseteq V$ with $s=|S|$ be a subset with
    \begin{itemize}
        \setlength{\itemsep}{0pt}
        \setlength{\parskip}{0pt}
        \item $3 \leq s \leq C_\alpha \frac{n}{d}$,
        \item $|E(S,S)| = o(s \log n)$, and
        \item $\sum_{v\in S} \deg(v) \geq s \frac{d}{100}$.
    \end{itemize}
    The last two conditions follow from
    \lemrefss{sparserandomgraph:propexpvertex:help1}
             {randomgraph:propsmalldist}
             {sparserandomgraph:propexpvertex:help2}.
    We show that $|\Gamma(S) \setminus S| >  C_\beta d s$
    with $C_\beta =\min\{1/200, e^{-500}/C_\alpha\}$.

        We may assume that all $\sum_{v \in S} \deg(v) - o(s \log n)$
    outgoing edges from $S$
    hit a uniformly chosen vertex among $V \setminus S$.
    This is a valid assumption as it may only lead to an
    underestimation of the number of outgoing edges
    since a vertex in $S$ may actually only hit the same vertex once.
    We call a set $S$ of size $s$ \emph{bad} if $|\Gamma(S) \setminus S| \leq  C_\beta d s$.
    We compute
    \begin{align*}
        \Pr{ \exists \, \mbox{bad set $S$ with $|S|=s$} }
        &\leq \binom{n}{s}
              \binom{n-s}{C_\beta d s}
              \left( \frac{C_\beta d s}{n}  \right)^{\sum_{v \in S} \deg(v) - o(s \log n)} \\
        &\leq \left( \frac{en}{s} \right)^s
              \left( \frac{e n}{C_\beta d s} \right)^{C_\beta d s }
              \left( \frac{C_\beta d s}{n}  \right)^{ds/110} \\
        &= \left( \frac{en}{s} \right)^s
              e^{C_\beta\,ds}
              \left( \frac{C_\beta d s}{n}  \right)^{(\frac{1}{110}-C_\beta)\,ds}\\
        &\leq \left( \frac{en}{s} \right)^s
              e^{C_\beta\,ds}
              \left( \frac{C_\beta d s}{n}  \right)^{ds/11000}
              \left( \frac{C_\beta d s}{n}  \right)^{ds/250}.
    \end{align*}
    Plugging in the definition of $s$ and $C_\beta$,
    we observe that the two middle terms of the last expression
    can together be upper-bounded by~$1$ since
    \begin{align*}
        e^{11000\,C_\beta} \, \left(\frac{C_\beta d s}{n}\right)
        \leq e^{11000\,C_\beta} \, C_\beta C_\alpha
        \leq e^{11000/200} \, e^{-500}
        = e^{-445}
        < 1.
    \end{align*}
    Hence,
    \begin{align*}
        \Pr{ \exists \, \mbox{bad set $S$ with $|S|=s$} }
        &\leq \left( \frac{en}{s} \right)^s
              \left( \frac{C_\beta d s}{n}  \right)^{ds/250} \\
        &= \exp \left( - s \left( \frac{d}{250} \,
              \log \left( \frac{n}{C_\beta ds}   \right) -
              \log \left( \frac{en}{s} \right) \right) \right) \\
        &\leq \exp \left( - 3 \left( \frac{\log n}{250} \,
              \log \left( \frac{1}{C_\alpha\,C_\beta}   \right) -
              \log \left( \frac{en}{3} \right) \right) \right)\\
        &\leq n^{-3},
       \end{align*}
       where the second last inequality holds due to our assumptions on $s$, $d\geq\log n$ and $C_\beta\leq e^{-500}/C_\alpha$.
       A union bound over
       all values for $s$ proves the claim
       of \lemref{sparserandomgraph:propexpvertex}.
\end{proof}
This proves that
sparse and dense random graphs satisfy all three properties of \expanding graphs
with probability \mbox{$1-o(1)$} and
therefore also
completes the proof of \thmref{randomgraph}.
\end{proof}

\subsubsection{Strong Expander Graphs}
\label{sec:expander}

Expander graphs (see \citet{HLW06} for a survey) are ``perfect'' networks
in the sense that they unite several desirable properties, such as low diameter,
small degree and high connectivity. They are therefore attractive for routing~\cite{BFSU94},
load balancing~\cite{RSW98} and communication problems such as the rumor spreading task considered here.

In order to define a strong expander graph more formally, we have to
introduce a bit of notation. For a $d$-regular graph $G$, its adjacency
matrix $A$ is symmetric and has $n$ real eigenvalues $d = \lambda_1 \geq
\lambda_2 \geq \cdots \geq \lambda_n$.
Define $\lambda := \max \left\{ | \lambda_2|, |\lambda_n| \right \}$. It
is well-known that $\lambda$ captures the expansion of $G$ in the sense
that
a small $\lambda$ implies good expansion (cf.\
\lemrefs{tanner}{expandermixing}) and vice
versa \cite[Theorem~2.4]{HLW06}.

\begin{definition}[expander]  \label{def:expander}
     We call a $d$-regular graph $G=(V,E)$ a \emph{strong expander}
    if there is a constant $C > 0$ (independent of $d$) such that $C < \sqrt{d}$ and $\lambda(G) \leq C \sqrt{d}$.
\end{definition}

We remark that graphs that satisfy the even stronger condition $\lambda \leq 2 \sqrt{d-1}$ are called \emph{Ramanujan graphs} and the construction of such graphs has received a lot of attention
(cf.~\citet{HLW06} for more details).
It is known that for any $d$-regular graph, $\lambda \geq 2 \sqrt{d-1} - \frac{2 \sqrt{d-1}-1}{nd/2}$. Hence as $n \rightarrow \infty$, the smallest possible value for the constant $C$ in Definition~\ref{def:expander} is $2 \sqrt{ (d-1)/d }$, in particular, we may assume in the following that $C > 1$.

We prove the following theorem, which has been used in \cite{CEOR12}.
\begin{thm}
\label{thm:expander}
    Strong expanders are \expanding.
\end{thm}

We first state two auxiliary lemmas that relate the second
largest eigenvalue in absolute value $\lambda$ to the expansion of $G$.
\begin{lem}[from~\cite{K95,T84}]\label{lem:tanner}
    For any subset $S \subseteq V$ of a $d$-regular graph $G$,
    \[
        |\Gamma(S)| \geq \frac{d^2 \, |S|}{\lambda^2 + (d^2 - \lambda^2) \, |S|/n}.
    \]
\end{lem}

\noindent
We also need the expander mixing lemma.
\begin{lem}[Expander Mixing Lemma, {\cite[Lemma~2.5]{HLW06}}]\label{lem:expandermixing}
    For any two subsets $A, B \subseteq V$ of a $d$-regular graph $G$, we have
    \[
        \left| |E(A,B)| - \frac{d |A| \cdot |B|}{n} \right| \leq \lambda \cdot \sqrt{ |A| \cdot |B|}.
    \]
\end{lem}

\noindent
We are now ready to prove \thmref{expander} that strong expanders are expanding.

\begin{proof}[Proof of \thmref{expander}]
        \propdeg\ is trivially satisfied as the graph is regular.
    We first prove \propexpvertex\ and afterwards \propexpedge.

    \noindent
    \propexpvertex:
    Let $S \subseteq V$ be any set of size $s=|S|\leq C_\alpha \frac{n}{d}$, where
    $C_\alpha\leq d/2$ is an arbitrary constant.
    Consider first the case $d=\omega(1)$.
    Then using \lemref{tanner} and $\lambda \leq C \sqrt{d}$ gives
    \begin{align*}
        |\Gamma(S)|
        \geq \frac{d^2 \, s}{\lambda^2 + (d^2 - \lambda^2) \, s/n}
        \geq \frac{d^2 \, s}{C^2 d + d^2 \frac{C_\alpha}{d}}
        = \frac{d s}{C^2 + C_\alpha}
    \end{align*}
    and therefore
    \[
        |\Gamma(S)\setminus S|
        \geq \left(\frac{1}{C^2 + C_\alpha} - \frac1d\right)\,d s.
    \]
    This proves \propexpvertex, as the factor in front of $ds$ is at least a constant (since $d=\omega(1)$).

    For $d=\Oh(1)$, we use \lemref{tanner} slightly differently to get
    \begin{align*}
     |\Gamma(S)| &\geq  \frac{d^2 \, s}{\lambda^2 + (d^2 - \lambda^2) \, s/n} \\
     &= \frac{d^2 s}{\lambda^2 \, (1-(s/n)) + d^2 \, (s/n) } \\
     &\geq \frac{d^2 s}{C^2 d \, (1-(s/n)) + d^2 \, (s/n) }.
    \end{align*}
    Hence,
    \begin{align*}
        |\Gamma(S)\setminus S|
        &\geq \frac{d^2 s}{C^2 d \, (1-(s/n)) + d^2 \, (s/n) } - s \\
        &= \frac{d-C^2 \, (1-(s/n)) - d \, (s/n) }{C^2 d \, (1-(s/n)) + d^2 \, (s/n)} \cdot ds.
    \end{align*}
    The denominator is bounded above by a constant, since $d=\Oh(1)$ and $s \leq n/2$. The numerator is at least a constant, since by assumption $C$ is a constant that is strictly smaller than $\sqrt{d}$. This proves \propexpvertex.

    \noindent
    \propexpedge:
    We may assume that $|S^c| \geq \lceil \frac{4n^2 C^2}{ds} \rceil$,
    as otherwise $|S^c| = \Oh( \frac{n^2}{ds} )$, and \propexpedge\ holds trivially by choosing the constant $C_{\omega}$ sufficiently large, for instance, $C_{\omega} := 10 \cdot \max\{C^2,1 \}$. Let us now order the vertices in $S^c$ according to the number of neighbors in $S$ in decreasing order. Let $N^{-}$ be the last $\lceil \frac{4n^2 C^2}{ds} \rceil$ vertices in that list, i.e., the $\lceil \frac{4n^2 C^2}{ds} \rceil$ vertices with the least number of neighbors in $S$ and let $N^{+}:=S^c \setminus N^{-}$ be the remaining set of vertices in $S^c$. Observing that $\lceil \frac{4n^2 C^2}{ds} \rceil \leq \frac{3}{2} \cdot \frac{4 n^2 C^2}{ds}$ (since $ds \leq n^2$ and $C \geq 1$) and applying \lemref{expandermixing}, we obtain
    \begin{align*}
       |E(S,N^{-})|
        &\geq d \, \frac{|S| \, |N^{-}|}{n} - \lambda \sqrt{|S| \, |N^-|} \\
         &\geq d \, \frac{s \,  \frac{4 n^2 C^2}{ds}}{n} -
          C \sqrt{d} \, \sqrt{s \, \frac{3}{2} \cdot \frac{4 n^2 C^2}{ds}} \\
         &= 4 C^2  n -  \sqrt{6} \cdot C^2  n \geq C^2 n.
    \end{align*}

     This implies that the average number of neighbors in $S$ of vertices in
     $N^{-}$ is at least
     \[
         \frac{ C^2 n}{\gamma \frac{\frac{3}{2} \cdot 4 n^2 C^2}{ds}}
         \geq \frac{ds}{6n}.
     \]
     and all vertices $N^{+}$ must have at least this degree.
     Hence we have shown that for every subset $S$, at least
     $|S^c| - |N^{-}| \geq n - s - \frac{3}{2} \cdot \gamma \frac{4 n^2 C^2}{ds}
     \geq n - s - 6 \frac{n^2 C^2}{ds} $
     vertices
     in $S^c$ have at least $ds/(6n)$ neighbors in $S$ and property \propexpedge\ follows with $C_{\delta}=1/6$ and $C_{\omega}=6 C^2 > 0$.
\end{proof}

\subsubsection{Random Graphs with Fixed Degree Sequence}
\label{sec:randomreg}

\begin{definition}[random graph with fixed degree sequence]
\label{def:randomreg}
Let $d_1, d_2, \ldots, d_n$ be a degree sequence with maximum degree
$\Delta=o(\sqrt{n}\,)$ and $\Delta/\delta=\Oh(1)$.
Then a random graph with this degree sequence
is chosen uniformly at random from the set of all
simple graphs with this degree sequence.
\end{definition}

Note that a random $d'$-regular graph
is a random graph with fixed degree sequence
$d_1=d_2=\cdots=d_n=d'$.
Random regular graphs
have gained increasing interest in the context of
peer-to-peer networks, e.g., they appear quite naturally as a limiting
distribution of certain graph transformations~\cite{MS06b,CDG07}.

For a random graph with fixed degree sequence as defined above,
\citet[Lemma~18]{BroderFSU98} showed that
$\lambda=\Oh(\sqrt{d}\,)$ with probability $1-\Oh(n^{-\poly(n)})$
and hence gave the following theorem.

\begin{thm}
\label{thm:randomreg}
    A random graph with fixed degree sequence is \expanding with probability \mbox{$1-o(1)$}.
\end{thm}

\subsection{Hypercubes}

We now recall the definition of hypercubes.
\begin{definition}[Hypercube]
    For any $d$, a $d$-dimensional hypercube $H=(V,E)$ has $n=2^d$ vertices
    $V=\{0,1\}^d$ and edges
    $E = \{ \{u,v\} \colon \| u-v \|_{1} = 1 \}$.
\end{definition}
The $i$-th bit of a bitstring $x \in \{0,1\}^d$ will be denoted as $x[i]$.
We observe that the hypercube is not expanding.
\begin{thm}
    \label{thm:HCnotexpanding}
    Hypercubes are not \expanding.
\end{thm}
\begin{proof}
    Define $S:=\bigcup_{i=1}^{\log d} L_i$, where $L_i$ is the set of vertices $x$ with
    $\|x\|_1 = i$.
    Then
    $3\leq |S|=o(n / \log n)$ and
    \[
        | \Gamma(S) \setminus S|
        = |L_{\log(d)+1}|
        = \binom{d}{\log(d)+1}
        = \frac{d-\log{d}}{\log(d)+1} \, \binom{d}{\log d}
        \leq \frac{d}{\log(d)+1} \,|S|
        = o(d \,|S|),
    \]
    which violates \propexpvertex.
\end{proof}
Hence a separate analysis is needed, and this is given in \secref{hypercube}.

\subsection{$k$-ary Trees}\label{sec:trees}

For complete \kary trees ($k\geq 2$) it is easy to verify
that they are not \expanding.

\begin{lem}
    \kary trees are not \expanding.
\end{lem}
\begin{proof}
Consider a \kary tree and let $C_{\alpha}=1/2$ and $S$ be the set of vertices which are in the subtree of a fixed children of the root. Then $|S| \leq (n-1)/k \leq n/2 \leq C_{\alpha} (n/d)$, but $|\Gamma(S) \setminus S|=1$ violating \propexpvertex.
\end{proof}

However, it is also not difficult to show the following theorem.
\begin{thm}\label{thm:tree}
    For complete \kary trees, the broadcast time of the quasirandom model is
    $\Oh(k \log (n)/\log k)$ with probability 1, while the expected broadcast time of the fully random model is $\Omega(k \log n)$.
\end{thm}
\begin{proof}
    As a \kary tree has a diameter of $\Theta( \log (n)/ \log k)$ and
    maximum degree of $k+1$, plugging these values into the bound of
    \thmref{degdiambound}, we obtain the first claim.

    To see the lower bound for the fully random model, define a path $P$ of length
    $\diam(G)/2$ inductively as follows.
    Assume that the root $u_0$ is initially informed. Then let
     $P=(u_0,u_1,\ldots,u_i)$ for $1 \leq i \leq
    \diam(G)/2$, where $u_{i}$ is the vertex which is the last one informed by~$u_{i-1}$.
    By the coupon collector's problem, the expected time it takes for $u_{i-1}$ to inform $u_{i}$ is at
    least $k \log k$ and therefore, the expected time to inform $v_{\diam(G)/2-1}$ is
    at least $\Omega(\diam(G) \, k \log k)=\Omega(k \log n)$.
\end{proof}

\section{Quasirandom Rumor Spreading on Expanding Graphs}
\label{sec:analysis}

In this section, we prove our main result that quasirandom rumor spreading informs all vertices in an expanding graph in a logarithmic number of rounds.
\begin{thm}
    \label{thm:main}
    Let $\gamma\geq1$ be a constant.
    The broadcast time of the quasirandom model on \expanding graphs
    is $\Oh(\log n)$ with probability $1-\Oh(n^{-\gamma})$.
\end{thm}

To analyze the propagation process, we decompose it into a forward part
(\secrefs{forward}{forwardlemmas})
and a backward part
(\secrefs{backward}{backwardlemmas}).
In the analysis of the forward part, we show that if a vertex is informed at some time, then $\Oh(\log n)$ steps
later,
only $\Oh(n/d)$ vertices remain uninformed (cf.~\thmref{forward}).
In the analysis of the backward part, we show that if a vertex is uninformed at some time, then $\Oh(\log n)$ steps
earlier,
at least $\omega(n/d)$ vertices must be uninformed as well (cf.~\thmref{backward}).
Combining both yields \thmref{main}.

We show that all this holds with probability $1-n^{-\gamma}$ for an
arbitrary $\gamma\geq1$.
As \thmref{main} is considerably easier to show for $d=\Oh(1)$, we handle this case separately
in \secref{const} and now concentrate on the case $d=\omega(1)$.  This makes the
proofs of the lemmas of this section slightly shorter.
Therefore in this section, apart from the last subsection, we may use the following adjusted property:
\begin{description}
    \item[\propdegg]
        $d=\omega(1)$ and $d=\Omega(\Delta)$.
        If $d=\omega(\log n)$ then $d=\Oh(\delta)$.
\end{description}

As the precise constants will be crucial in parts of the following proofs,
we use the following notation.  Constants with a lowercase Greek letter index
(e.g., $C_\alpha$ and~$C_\beta$) stem from \defref{ourgraph}.  Constants without
an index or with a numbered index (e.g.,
$C$ and $C_1$) are local constants in lemmas.  $K$ is used to denote
a number of time steps.

\subsection{Forward Analysis}
\label{sec:forward}

In this section we prove the following theorem.

\begin{thm}
    \label{thm:forward}
    Let $\gamma\geq1$ be a constant. The probability that the quasirandom model
    started in a fixed vertex $u$ informs $n-\Oh(n/d)$ vertices
    within $\Oh(\log n)$ rounds is at least $1-n^{-\gamma}$.
\end{thm}

In our analysis we use the following two notations for sets of informed vertices.
Let $I_t$ be the set of vertices that know the rumor after the $t$-th step.
Let $N_t\subseteq I_t$ be the set of ``newly informed'' vertices, that is, those which
know the rumor after the $t$-th step, but have not spread this information yet.
The latter set will be especially important as these are the vertices which
have preserved their independent random choice.

Each of the following Lemmas~\ref{lem:forwardwarmup}--\ref{lem:forwardfinal}
examines one phase consisting of several steps. Within each phase,
we will only consider information spread from vertices
that became informed in the previous phase.  This is justified by \lemref{delay}.

Let $u$ be (newly) informed at time step $0$.  To get
a sufficiently large set of newly informed vertices to start with,
we first show how to obtain a set $N_t$ of size $\Theta(\log n)$ within
$t=\Oh(\log n)$ steps.
This is simple if $d=\omega(\log n)$---after $c \log n$ rounds, the first vertex has informed exactly $c \log n$ new vertices.
Otherwise, we use the fact that \propexpvertex\ implies that the neighborhoods $\Gamma^k(u)$
grow exponentially with $k$.  Since within $\Delta$ steps, $\Gamma^k(u)$ becomes
informed if $\Gamma^{k-1}(u)$ was informed beforehand, this yields the claim in this
case.
The precise statement is as follows.

\begin{lem}\label{lem:forwardwarmup}
    Let $C > 0$ be an arbitrary constant.
    Then with probability~1 there is a time step $t = \Oh(\log n)$ such that
    \begin{itemize}
    \setlength{\itemsep}{0pt}
    \setlength{\parskip}{0pt}
    \item $|N_t| \geq C \log n$ and
    \item $|I_t \setminus N_t| = o(|N_t|).$
    \end{itemize}
\end{lem}

\noindent
The proof of \lemref{forwardwarmup} and all following lemmas
can be found in \secref{forwardlemmas}.
We now assume that we have a set $N_t$ of size $\Omega(\log n)$.
We aim at informing $\Omega(n/d)$ vertices.
For the very dense case of $d=\Omega(n/\log n)$ this is a trivial statement.
Note that in the following argument we can always assume that we have not informed
\emph{too many} vertices as the number of informed vertices can
at most double in each time step.
The following lemma shows that given a set of informed vertices
matching the conditions of \propexpvertex, within a constant number
of steps the set of informed vertices increases by a factor strictly
larger than one.

\begin{lem}\label{lem:forwardstepfirst}
    For any constants $\gamma\geq1$ and $C_\alpha>0$
    there are constants
    $K\geq1$,
    $C_1>1$,
    $C_2>1$, and
    $C_3\in(3/4,1)$
    such that for all time steps~$t$, if
    \begin{itemize}
    \setlength{\itemsep}{0pt}
    \setlength{\parskip}{0pt}
    \item $C_1\log n \leq |I_t| \leq C_\alpha\,(n/d)$ and
    \item $|N_t| \geq C_3\, |I_t|$,
    \end{itemize}
    then with probability $1-n^{- \gamma}$,
    \begin{itemize}
    \setlength{\itemsep}{0pt}
    \setlength{\parskip}{0pt}
    \item $|I_{t+K}| \geq C_2\, |I_t|$ and
    \item $|N_{t+K}| \geq C_3\, |I_{t+K}|$.
    \end{itemize}
\end{lem}

\noindent
As the precondition of the next \lemref{forwardstepsecond} is
$|I_t| \geq 16\,C_\omega (n/d)$, let $C_\alpha= 16\,C_\omega$. Then \lemref{forwardstepfirst} yields a
constant $C_2 > 1$ such that applying this lemma
at most $\log_{C_2}\big(16\,C_\omega\,(n/d)\big)=\Oh(\log n)$
times leads to at least
$16\,C_\omega (n/d)$ informed vertices, a constant fraction of which is newly informed.

The next aim is informing a linear number of vertices.
Note that as long as that is not achieved, \propexpedge\ implies
that there is a large set of uninformed
vertices which have many neighbors in $N_t$.
This is the main ingredient of the following \lemref{forwardstepsecond}.
It shows that under these conditions, a phase of a constant number of steps
suffices to triple the number of informed vertices.

\begin{lem}\label{lem:forwardstepsecond}
    For any constant $\gamma\geq1$
    there are constants
    $K\geq1$,
    $C>1$,
    and
    $C_\omega>0$
    such that for all time steps $t$, if
    \begin{itemize}
        \setlength{\itemsep}{0pt}
        \setlength{\parskip}{0pt}
        \item $\max \{ C \log n, 16\,C_\omega (n/d) \} \leq |I_t| \leq n/16$ and
        \item $|N_t| \geq (3/4) \, |I_t|$,
    \end{itemize}
    then with probability $1 - n^{-\gamma}$,
    \begin{itemize}
        \setlength{\itemsep}{0pt}
        \setlength{\parskip}{0pt}
        \item $|I_{t+K}| \geq 3 \, |I_t|$ and
        \item $|N_{t+K}| \geq (3/4) \, |I_{t+K}|$.
    \end{itemize}
\end{lem}

\noindent
Applying \lemref{forwardstepsecond} at most $\Oh(\log n)$ times,
a linear fraction of the vertices gets informed.
In a final phase of $\Oh(\log n)$ steps, one can then inform all
but $\Oh(n/d)$ vertices as shown in the following \lemref{forwardfinal}.

\begin{lem}\label{lem:forwardfinal}
    For any constants $\gamma\geq1$ and $C > 0$ there is a $K=\Oh(\log n)$ such that for all time steps $t$, if
    \begin{itemize}
        \setlength{\itemsep}{0pt}
        \setlength{\parskip}{0pt}
        \item
         $|N_t| \geq C \, n$,
    \end{itemize}
    then with probability $1 - n^{-\gamma}$,
    \begin{itemize}
        \setlength{\itemsep}{0pt}
        \setlength{\parskip}{0pt}
        \item
         $|I_{t+K}| = n - \Oh(n/d)$.
    \end{itemize}
\end{lem}

Combining all above phases, a union bound gives
$|I_{\Oh(\log n)}| = n - \Oh(n/d)$
with probability $1 - \Oh(\log (n)\,n^{-\gamma})$.
As $\gamma$ was arbitrary in all lemmas, \thmref{forward} follows.

\subsection{Proofs of the Lemmas Used in the Forward Analysis}
\label{sec:forwardlemmas}

\begin{proof}[Proof of \lemref{forwardwarmup}]
    Let $u$ be informed at time step $0$. If $d =\omega(\log n)$,
    then by \propdeg\ $\delta=\Theta(d)$ and a single phase of $C \log n$ rounds suffices, that is, we have $N_{C \log n}=C \log n$,
    and the lemma follows.

    We now describe how to obtain $C \log n$ newly informed vertices for $d = \Oh(\log n)$.
    For this, we choose a $C_\alpha$ such that $C_\alpha n/d \geq C\log n$
    and get, by \propexpvertex\ for $k \geq 3$,
    as long as $| \Gamma^{\leq k}(v) |=\Oh(n/d)$,
    \begin{align}
        | \Gamma^{\leq k+1 }(v) |
        &= | \Gamma^{\leq k }(v) | + | \Gamma^{k+1 }(v) |
        = | \Gamma^{\leq k }(v) | + | \Gamma(\Gamma^{\leq k}(v))\setminus \Gamma^{\leq k}(v) |\notag\\
        &\geq (1+ C_\beta d)\, |\Gamma^{\leq k}(v)|.
         \label{eq:diameterbound:Gamma}
    \end{align}
    Subtracting $| \Gamma^{\leq k}(v) |$ on both sides yields
    \begin{equation*}
    | \Gamma^{k+1 }(v) | \geq C_{\beta} \, d \, |\Gamma^{\leq k}(v)|.
    \end{equation*}
    As $|\Gamma^{\leq 3}(v)| \geq 3$,  by induction,
    \begin{equation*}
       | \Gamma^{ k }(v) |
        \geq 3 \, (C_\beta \, d)^{k-3}
    \end{equation*}
    for all $k$ with $k \geq 3$ and $|\Gamma^{\leq k-1}(v)| \leq C \log n$.
    Therefore we can choose a $k=\Oh(\log \log (n)/\log d)$ such that $| \Gamma^{k}(v)| \geq C \log n$.

    We use the delaying and ignoring assumption (cf.~Lemma~\ref{lem:delay}) to perform $k$ phases of $\Delta$ rounds each. Then after these $t=\Delta k=\Oh( \Delta \,(\log\log n) / \log d)=\Oh(\log n)$
    steps
    (as $\Delta=\Oh(d)$ by \propdeg\ and
        $d/\log d=\Oh(\log(n) / \log\log n)$ by $d=\Oh(\log n)$)
    all vertices in $\Gamma^{\leq k}(v)$ get informed, but no vertex of $\Gamma^{k}(v)$ has been active. In consequence,
    we have
    \begin{align}
        |N_t| &=
            |\Gamma^k(v)| \geq C\log n,
            \label{eq:forwardwarmup:Gamma}
            \\
        |I_t\setminus N_t| &=
            |\Gamma^{\leq k-1}(v)| \leq |\Gamma^k(v)| / (C_\beta \, d) = o(|N_t|)
            \notag ,
    \end{align}
    where the last equation stems from \propdegg.
\end{proof}

\begin{proof}[Proof of \lemref{forwardstepfirst}]
    We choose the following constants:
    \begin{align*}
        C_1&:=\tfrac{8 \,\gamma \,\Delta^2}{C_\beta^2 \,d^2}>1,
        &C_2:=\tfrac{4 \,\Delta}{C_\beta \,d}>1,\\
        C_3&:=\left(1 - \tfrac{C_\beta \,d}{4\,\Delta}\right)\in(3/4,1),
        &K:=\big\lceil\big(\tfrac{3\,\Delta}{C_\beta \,d}\big)^2\big\rceil\geq1,
    \end{align*}
    where the $C_\beta$ is from \propexpvertex\ and depends on the given $C_\alpha$.
    $K$ and $C_1$ to $C_3$ are all $\Theta(1)$ by \propdeg.
    As $I_t$ is a connected set of appropriate size, \propexpvertex\ gives
    \begin{equation}
        \label{eq:forward:step:1}
        |\Gamma(I_t) \setminus I_t| \geq  C_{\beta} \, d \, |I_t|.
    \end{equation}
    Since we are interested in the expansion of $N_t$ and not of $I_t$,
    we calculate
    \begin{align}
        | \Gamma(I_t) \setminus I_t |
        &= \big| \big(\Gamma(I_t \setminus N_t) \setminus I_t\big)
             \cup \big(\Gamma(N_t) \setminus I_t\big) \big| \notag\\
        &\leq \big| \Gamma(I_t \setminus N_t) \setminus I_t \big|
             + \big| \Gamma(N_t) \setminus I_t \big| \notag\\
        &\leq \Delta | I_t \setminus N_t |
             + \big| \Gamma(N_t) \setminus I_t \big|.
             \label{eq:forward:step:2}
    \end{align}

    \noindent
    Combining \eqs{forward:step:1}{forward:step:2} with the
    assumption $|I_t \setminus N_t| \leq \frac{C_\beta \,d}{4\,\Delta} |I_t|$,
    \begin{equation*}
                \big| \Gamma(N_t) \setminus I_t \big|
            \geq C_{\beta} \, d \, |I_t| - \Delta | I_t \setminus N_t |
            \geq 3\,C_{\beta} \, d \, |I_t| / 4.
    \end{equation*}

    \noindent
    We now perform one phase consisting of $K$ rounds. We compute the size of the resulting sets $I_{t+K}$ and $N_{t+K}$ as follows.

    Let $v \in \Gamma(N_t) \setminus I_t$.
    Then there is a $u \in N_t$ such that $(u,v) \in E$.
    The probability that $u$ contacts $v$ within this time interval is $\min\{K/\deg(u),1\} \ge
    K/\Delta$ (as $\Delta=\omega(1)$ by \propdegg), which naturally is a lower bound for $v$ becoming contacted by an
    arbitrary vertex of $N_t$.
    By linearity of expectation,
    the expected number of vertices becoming contacted is at least
    \begin{align*}
       \Ex{| N_{t+K} |} &\geq K \,|\Gamma(N_t) \setminus I_t|/\Delta
        \geq  3\,C_{\beta} K \,d \, |I_t| / (4 \Delta).
    \end{align*}

    \noindent
    As every vertex can only contact at most $K$ vertices in this
    time interval, Azuma's inequality (cf.~\lemref{Azuma}) gives a probabilistic lower bound
    on the number of newly informed vertices.
    More precisely,
    \begin{align*}
        \Pr{ | N_{t+K} | \leq \frac{C_{\beta} \,K \,d \, |I_t| }{ 2 \Delta} }
        &\leq
        \exp\left(-
        \frac{ C_\beta^2 \,d^2\, |I_t|^2 }
             { 8\, \Delta^2 \,|N_t|  }
             \right)
        \leq
        n^{- C_1\, C_\beta^2 \,d^2 / (8 \,\Delta^2) }
        =
        n^{-\gamma }.
    \end{align*}

    \noindent
    It remains to check that $| N_{t+K} | \geq \frac{C_{\beta} \,K \,d \, |I_t| }{ 2 \,\Delta}$ implies the two parts of the claim.
    First,
    \begin{align*}
        |I_{t+K}|
        \geq |N_{t+K}|
        \geq \frac{C_{\beta} \,K \,d}{ 2 \,\Delta} \, |I_t|
        \geq \frac{4 \,\Delta}{C_\beta \,d} \, |I_t|
        = C_2 \, |I_t|.
    \end{align*}
        For the second part, observe that
    \begin{align*}
        |N_{t+K}|
        \geq \frac{C_{\beta} \,K \,d \, |I_t| }{ 2 \,\Delta}
        \geq \frac{C_{\beta} \,K \,d \, (|I_{t+K}|-|N_{t+K}|) }{ 2 \,\Delta}
        = \frac{C_{\beta} \,K \,d}{ 2 \,\Delta} \, |I_{t+K}|
             - \frac{C_{\beta} \,K \,d}{ 2 \,\Delta} \, |N_{t+K}|.
    \end{align*}
    Rearranging  yields
    \begin{align*}
        |N_{t+K}|
        &\geq \frac{C_{\beta} \,K \,d}{ 2 \,\Delta + C_{\beta} \,K \,d} \, |I_{t+K}|
        \geq \frac{C_{\beta} \big(\tfrac{3\,\Delta}{C_\beta \,d}\big)^2 \,d}
          { 2 \Delta + C_{\beta} \big(\tfrac{3\,\Delta}{C_\beta \,d}\big)^2 \,d}
          \, |I_{t+K}| \\
        &= \frac{9 \Delta}{ 2\, C_\beta \,d + 9\, \Delta } \, |I_{t+K}|
        \geq \left(1 - \frac{C_\beta \,d}{4\,\Delta}\right) |I_{t+K}|.
        \qedhere
    \end{align*}
\end{proof}

\begin{proof}[Proof of \lemref{forwardstepsecond}]
    We choose $C:=\frac{512\, \gamma^3\,\Delta^2}{3\,C_\delta^2\, d^2}>1$,
    $K:= \big\lceil\frac{16\,\gamma\,\Delta}{C_\delta \,d}\big\rceil\geq1$,
    and $C_\omega>0$ according to \propexpedge.

    By property \propexpedge, the number of vertices in $N_t^c$ which have
    at least $C_\delta d \, |N_t| /n$ neighbors in $N_t$ is at least
    $|N_t^c|-\frac{C_\omega \,n^2}{d\,|N_t|}$.
    Therefore,
    the number of vertices in $I_t^c$ which have at least
    $C_\delta d(|N_t| /n )$ neighbors in $N_t$ is at least
    \begin{align*}
        |N_t^c|-|I_t|-\tfrac{C_\omega \,n^2}{d\,|N_t|}
        \geq
        n-2\,|I_t|-n/12
        \geq
        19n/24\geq
        3n/4,
    \end{align*}
    where the first inequality is due to $16\,C_\omega (n/d) \leq |I_t| \leq 4/3 \,|N_t|$.

    \noindent
    We call a vertex $v\in I_t^c$ \emph{good} if it has
    at least $C_\delta d\, |N_t| /n $ neighbors in $N_t$.
    The probability that a good vertex gets informed in
    a phase of $K$ rounds
    (again using $K\leq\Delta=\omega(1)$ by \propdegg)
    is at least
    \begin{align*}
        1- \left( 1 - \frac{K}{\Delta} \right)^{C_\delta d\,|N_t| /n}
        &\geq
        1 -\exp \big( -\tfrac{K C_\delta d\,|N_t| }{\Delta n}\big)
        \geq
        1 -\exp ( -16\,\gamma\,|N_t|/n)\\
        &\geq
        1 -\tfrac1{(16\,\gamma\,|N_t|/n) +1}
        =
        \tfrac{16\,\gamma\,|N_t|}{16\, \gamma\,|N_t| +n}.
    \end{align*}

    \noindent
    By linearity of expectation,
    \begin{align*}
        \Ex{|N_{t+K}|}
        \geq
        \tfrac{16\,\gamma\,|N_t|}{16\,\gamma\,|N_t| +n}
        \tfrac{3n}{4}
        \geq
        \tfrac{16\,\gamma\,|N_t|}{16\,\gamma\,n/16 +n}
        \tfrac{3n}{4}
        =
        \tfrac{\gamma\,|N_t|}{(\gamma/16) +1/16}
        \tfrac{3}{4}
        \geq
        6 \,|N_t|.
    \end{align*}
    Azuma's inequality (cf.~\lemref{Azuma}) gives
    \begin{align*}
        \Pr{ |N_{t+K}| \leq 4\,|N_t| }
        &\leq \exp\left( -\frac{2\,(2|N_t|)^2 }{ |N_t|\,K^2} \right)
        = \exp\left( -\frac{8 |N_t| }{ K^2} \right)\\
&\leq \exp\left( -\frac{|N_t| C_\delta^2 d^2 }{ 128 \,\gamma^2\Delta^2} \right)
\leq \exp\left( -\frac{3 C \log(n) C_\delta^2 d^2 }{ 512 \,\gamma^2\Delta^2} \right)
        = n^{-\gamma}.
    \end{align*}
    Therefore with probability $1-n^{-\gamma}$,
    \begin{align*}
        |N_{t+K}|
        &\geq 4\,|N_t|
        \geq 3 \,|I_t|
        = 3 \,|I_{t+K}| - 3 \,|N_{t+K}|
        \intertext{and after rearranging,}
        |N_{t+K}|
        &\geq \tfrac34 |I_{t+K}|.
    \end{align*}
    This proves the first claim.  The second claim follows from
    \[
        |I_{t+K}| \geq |N_{t+K}| \geq 4 |N_t| \geq 3\, |I_t|.
        \qedhere
    \]
\end{proof}

\begin{proof}[Proof of \lemref{forwardfinal}]
    Let $X \subseteq N_t^c$ be the set of vertices in $N_t^c$ that
    have at least $C_{\delta}\, d\, |N_t|/n$
    neighbors in $N_t$. By \propexpedge,
    \begin{align*}
     |X| &\geq (n-|N_t|) - \frac{C_{\omega} \,n^2}{d|N_t|}
     \geq n - |N_t| - \Theta\big(\tfrac{n}{d}\big).
    \end{align*}
    Let $v \in X$ and consider a phase of
    $K:=\big\lceil\tfrac{2\, \gamma\, \Delta \,n}{C_{\delta} \,|N_t| \,d} \log n\big\rceil$ rounds.
    Note that $K=\Oh(\log n)$ by \propdeg.

    If $K \geq \Delta$, $v$ becomes informed in this phase with probability $1$.
    Otherwise, the probability that $v$ will not be informed in this phase is at most
    \begin{align*}
     \Pr{ v \notin N_{t+K}}
     \leq \left( 1 - \frac{K}{\Delta} \right)^{C_{\delta} |N_t| d/n}
     \leq \exp( - 2 \,\gamma\, \log n) = n^{-2 \, \gamma}.
    \end{align*}
    Taking the union bound over all vertices in $X$,
    we obtain that all vertices in $X$ get informed with probability $1-n^{-\gamma}$.
    The claim follows.
\end{proof}

\subsection{Backward Analysis}
\label{sec:backward}

The forward analysis has shown that within $\Oh(\log n)$ steps, at most $\Oh(n/d)$ vertices
stay uninformed.  We now analyze the reverse.  The question here is how many vertices
have to be uninformed at time $t-\Oh(\log n)$ if there is an uninformed vertex at time $t$.
We will show that this is at least $\omega(n/d)$.
To formalize this, recall that $U_{[t_1,t_2]}(w)$ is the set of vertices that reach the vertex $w$ within the time interval $[t_1,t_2]$ (using the usual meaning of ``reach'' as defined on page~\pageref{def:reach}).
We will prove the following theorem.

\begin{thm}
    \label{thm:backward}
    Let $\gamma\geq1$ be a constant.
    If the quasirandom rumor spreading process does not
    inform a fixed vertex $w$ until some time $t$, then there are
    $\omega(n/d)$ uninformed vertices at
    time $t-\Oh(\log n)$ with probability at least $1-n^{-\gamma}$.
\end{thm}

To prove \thmref{backward}, we fix an arbitrary vertex $w$ and a time $t$.
Ignoring some technicalities, our aim is to prove a lower bound on the number of
vertices which have to be uninformed at times before $t$ to keep $w$ uninformed at time
$t$. We first show that the set of
uninformed vertices at time $t-\Oh(\log n)$
is at least of logarithmic size.

For $d=\Oh(\log n)$ this follows from \propexpvertex\ as all vertices of
$\Gamma^{\Oh(\log\log n/\log d)}(w)$ (and there are at least $\Omega(\log n)$
of these) reach $w$ within
$\Oh(\log n)$ steps.  For $d=\omega(\log n)$, a simple Chernoff bound shows
that enough vertices of $\Gamma(w)$ contact $w$ within $\Oh(\log n)$ steps.
This is summarized in the following lemma.  The proofs of all three
lemmas of this section can be found in the following \secref{backwardlemmas}.

\begin{lem}\label{lem:backwardwarmup}
    Let $\gamma\geq1$ and $C \geq 1$ be constants,
    $w$ a vertex,
    and $t_2=\Omega(\log n)$ a time step.
    Then with probability $1 - 2 \, n^{-\gamma}$ there is a time step
    $t_1 = t_2  - \Oh( \log n )$
    such that
    \[
        |U_{[t_1,t_2]}(w)| \geq C \, \log n.
    \]
\end{lem}

\noindent
We now know that within a logarithmic number of time steps, there are at least
$c\log n$ vertices which have reached $w$.  Very similarly to
\lemrefs{forwardstepfirst}{forwardstepsecond} in the forward analysis,
we can increase the set of vertices that reach $w$
by a multiplicative factor by going back a constant number of time steps.  The following lemma again mainly
uses~\propexpvertex.
For the very dense case of $d=\Omega(n/\log n)$, there is nothing to show.

\begin{lem}\label{lem:backwardstep}
    For any constant $\gamma\geq1$ there is a constant $K$ such that
    for all vertices~$w$ and time steps $t_1,t_2$, if
    \[
        \log n \leq |U_{[t_1,t_2]}(w)| = \Oh(n/d),
    \]
    then with probability $1 - n^{-\gamma}$,
    \[
        |U_{[t_1-K,t_2]}(w)| \geq 4\, |U_{[t_1,t_2]}(w)|.
    \]
\end{lem}

\noindent
Using \lemref{backwardstep} at most $\Oh(\log n)$ times, we obtain a set of vertices
that reach $w$ of size $\Omega(n/d)$.
If these are $\omega(n/d)$ vertices, we are done.
Otherwise, the following \lemref{backwardfinal} shows that
a phase consisting of $\Oh(\log n)$ steps suffices to get to this point.  This is the only lemma
which substantially uses \propdegg.

\begin{lem}\label{lem:backwardfinal}
    Let $\gamma\geq1$ be a constant,
    $w$ a vertex,
    and $t_1,t_2$ time steps such that
    \[
        |U_{[t_1,t_2]}(w)| = \Theta(n/d).
    \]
    Then with probability $1 - n^{-\gamma}$,
    \[
        |U_{[t_1-\Oh(\log n),t_2]}(w)| = \omega(n/d).
    \]
\end{lem}

\noindent
This finishes the backward analysis and shows that $\omega(n/d)$ vertices have to be uninformed
to keep a single vertex uninformed for $\Oh(\log n)$ steps.  Together with the forward
analysis, which proved that only $\Oh(n/d)$ vertices remain uninformed after $\Oh(\log n)$
steps, this finishes the proof of \thmref{main} for $d=\omega(1)$.

\subsection{Proofs of the Lemmas Used in the Backward Analysis}
\label{sec:backwardlemmas}

\begin{proof}[Proof of \lemref{backwardwarmup}]
    Consider first the case that $d = \Oh(\log n)$.
    In this case, we choose, as in the proof of
    \lemref{forwardwarmup},
    a constant $C_\alpha$ such that $C_\alpha n/d \geq C\log n$
    and apply \propexpvertex.
    By \eq{forwardwarmup:Gamma} from page~\pageref{eq:forwardwarmup:Gamma},
    there exists a $k=\Oh(\log \log(n)/ \log d)$ such that
    \begin{align*}
        |\Gamma^{\leq k}(w)| \geq |\Gamma^{k}(w)| \geq C \log n.
    \end{align*}
    Since within $\Delta$ rounds each vertex has contacted all neighbors, we
    have
    $
        \Gamma^{\leq i}(w) \subseteq U_{[t_2- i \Delta,t_2]}(w)
    $
    for $i\geq 1$
    and therefore
    $
    \Gamma^{\leq k}(w) \subseteq
    U_{[t_2- k \Delta,t_2]}(w).
    $
    As $k \Delta = \Oh(\log n)$, we see
    that $| U_{[t_2-\Oh(\log n),t_2]} | \geq C \log n$ with probability~1.

    In the remaining case $d=\omega(\log n)$
    we estimate the number of neighbors of $w$ which reach $w$ in the previous
    $K:=\lceil 4C^2\gamma\Delta\log(n)/\delta \rceil$ steps.
    Note that $K=\Oh(\log n)$ by \propdeg.
    For each neighbor $u \in \Gamma(w)$, define a random variable $X(u)$, which is one if $u$
    contacts~$v$
    within the time interval $[t_2 - K,t_2]$, and zero otherwise.
    Then for each $u \in \Gamma(w)$,
    $
       \Pr{ X(u)=1 }
       \geq K/\Delta.
    $
    We define $X:= \sum_{u \in \Gamma(w)} X_u$.
    Linearity of expectation gives
    $
      \Ex{X} \geq K\,\delta/\Delta \geq 4 C^2 \gamma \log n.
    $
    Since $\{ X(u) \colon u \in \Gamma(w) \}$ is a set of
    independent random variables, we obtain by a Chernoff bound that
    \begin{align*}
        \Pr{ X \leq C\,\log n } &\leq \Pr{ X \leq \tfrac{1}{4} \Ex{X} }
        \\ &\leq \exp \left( - (3/4)^2 \Ex{X} / 2  \right)
        \\ &= \exp \left( - (9/32)\, 4 C^2 \gamma \log n \right) \leq n^{-\gamma},
            \end{align*}
    where we used the assumption $C \geq 1$. This implies
    that with probability $1 - n^{-\gamma}$,
    we have
    \[
      \left|U_{[t_2 - \Oh(\log n),t_2]}(w)\right|
      \geq C \log n.
      \qedhere
    \]
\end{proof}

\begin{proof}[Proof of \lemref{backwardstep}]
    Let $S:=U_{[t_1,t_2]}(w)$ and let $|S| \leq C_\alpha \, (n/d)$
    for a constant $C_\alpha$.
    As $S$ is a connected set, \propexpvertex\ gives
    \[
        |\Gamma(S) \setminus S| \geq  C_{\beta} \, d \, |S|.
    \]

    \noindent
    for a suitable constant $C_{\beta}$.
    Let $K=\big\lceil\tfrac{8 \,\gamma}{C_\beta}
          \tfrac{\Delta}{d}\big\rceil=\Oh(1)$ (by \propdeg).
            As every vertex $u\in\Gamma(S)\setminus S$ has at least one edge to a vertex $v\in S$,
    the probability that a vertex $u\in\Gamma(S)\setminus S$ contacts a $v\in S$
    in the interval $[t_1-K,t_1-1]$ is at least $K/\Delta$ and
    $S':=U_{[t_1-K,t_2]}(w)$.
    By linearity of expectation,
    the expected number of vertices in $S' \setminus S$
    is at least
    \begin{align*}
       \Ex{ |S' \setminus S| } &\geq K |\Gamma(S)\setminus S|/\Delta
        \geq  C_{\beta} K d \, |S| / \Delta.
    \end{align*}

    \noindent
                A simple application of the Chernoff bound gives
    \begin{align*}
        \Pr{ |S' \setminus S| \leq \frac{C_{\beta} K d \, |S| }{ 2 \Delta} }
        \leq \exp\left(- \frac{C_{\beta} K d \, |S| }{ 8 \Delta} \right)
        \leq n^{- \frac{C_{\beta} K d}{ 8 \Delta}}.
    \end{align*}
                 Hence with probability $1-n^{- \gamma }$,
    \[
        |S'|
        \geq \frac{C_{\beta} K d \, |S| }{ 2 \Delta}
        \geq 4 \gamma |S|
        \geq 4\,|S|.
        \qedhere
    \]
\end{proof}

\begin{proof}[Proof of \lemref{backwardfinal}]
    Let $S:=U_{[t_1,t_2]}(w)$ with $|S| \leq C_\alpha \, (n/d)$
    for a constant $C_\alpha$.
    Also let
    $K:=\big\lceil
    \frac{8 \gamma}{C_\beta}
    \frac{\Delta}{d}
    \frac{n}{|S| \,d}
    \log n
    \big\rceil$ and
    $S':=U_{[t_1-K,t_2]}(w)$.
    Note that $K=\Oh(\log n)$ by \propdeg.
    We examine a phase of $K$ steps.

    As $S$ is a connected set, \propexpvertex\ gives, as in the proof
    of \lemref{backwardstep},
    $
        |\Gamma(S) \setminus S| \geq  C_{\beta} \, d \, |S|.
    $
    If $K\geq\Delta$, the lemma immediately follows from the observation
    \[
        |S'| = |\Gamma^{\leq1}(S)| = \Theta(d\,|S|) = \Theta(n) = \omega(n/d).
    \]
    The last equality is based on $d=\omega(1)$ as given by \propdegg.

    We now assume $K\leq\Delta$.
    As every vertex $u\in\Gamma(S)\setminus S$ has at least one edge to a vertex $v\in S$,
    the probability that a vertex $u\in\Gamma(S)\setminus S$ contacts a $v\in S$
    in the interval $[t_1-K,t_1-1]$ is at least $K/\Delta$.
    By linearity of expectation,
    the expected number of vertices in $S' \setminus S$
    is at least
    \[
        \frac{K}{\Delta} |\Gamma(S)\setminus S|
        \geq \frac{C_{\beta} K d \, |S|}{\Delta}
        \geq \frac{8 \gamma\,n \log n}{d}
    \]

    \noindent
                Again, a Chernoff bound gives
    \begin{align*}
        \Pr{ |S' \setminus S| \leq \frac{4 \gamma\,n \log n}{d} }
        \leq \exp\left(- \frac{\gamma\,n \log n}{d} \right)
        \leq n^{- \gamma }.
    \end{align*}

    \noindent
    Hence $|S'|\geq|S' \setminus S|= \Omega( n \log(n)/d)=\omega(n/d)$
    with probability $1-n^{- \gamma }$ for $K\leq\Delta$.
\end{proof}

\subsection{Analysis for Graphs with Constant Degree}
\label{sec:const}

It remains to show that the quasirandom model also
works well on \expanding graphs with constant degree $d=\Oh(1)$.
To do this, we apply \thmref{degdiambound} to see that for any graph the quasirandom model succeeds in $\Delta \cdot \diam(G)$ steps.
The corresponding bound for the fully random model is
$\Oh(\Delta \, (\diam(G) + \log n))$
with probability $1-n^{-1}$~\cite[Theorem~2.2]{FPRU90}.

Naturally, the diameter of \expanding graphs can be bounded easily as follows (cf.~\cite[p.~455]{HLW06} for a related result).
Plugging \lemref{diameterbound}
into the upper bound of $\Delta \cdot \diam(G)$ yields \thmref{main} for $d=\Oh(1)$.

\begin{lem}
    \label{lem:diameterbound}
    For any \expanding graph $G$ with $d=\Oh(1)$, $\diam(G) = \Oh(\log n)$.
\end{lem}
\begin{proof}
    Fix two vertices $v$ and $w$.
    We show that $\Gamma^{\leq\Oh(\log n)}(v)\cup\Gamma^{\leq\Oh(\log n)}(w)\neq\emptyset$.
    As $G$ is connected, $|\Gamma^{\leq 3}(v)|\geq 3$.
    Now we choose $C_\alpha=d/2$ (which is valid since $d$ is a constant)
    and proceed as in the proof of \lemref{forwardwarmup}.
    By \propexpvertex\ we again get \eq{diameterbound:Gamma}
    for $k > 3$,
    and therefore by induction
    \begin{equation*}
        |\Gamma^{\leq k}(v)|
        \geq 3\,(1+C_\beta d)^{k-3}
    \end{equation*}
    for all $k$ with $k>3$ and $|\Gamma^{\leq k-1}(v)| \leq n/2$.
    Therefore we can choose a $k$ such that $| \Gamma^{\leq k}(v)| \geq n/2$
    and $k=\Oh(\log n)$.
    As analogously $| \Gamma^{\Oh(\log n)}(w)| \geq n/2$,
    we can conclude that there is a path of length $\Oh(\log n)$ from
    $v$ to $w$.
\end{proof}

\section{Lower Bounds for the Fully Random Model on Sparse Random Graphs}\label{sec:lowerbounds}

In this section, we discuss lower bounds for the fully random model on sparse random graphs. They will show that the quasirandom model is superior on such graphs. \citet{FPRU90} proved the following bound.
\begin{thm}[{\cite[Theorem~4.1]{FPRU90}}]\label{thm:feige}
Let $p=(\log n + f(n))/n$, where $f(n)=\omega(1)$ and $f(n)=\Oh(\log \log n)$. Then for almost all random graphs $G(n,p)$, the broadcast time of the fully random model is $\Omega(\log^2 n)$ with probability at least $n^{-1}$.
\end{thm}

\thmref{feige} stems simply from the fact that with high probability such graphs contain a vertex having constant degree with all neighbors having logarithmic degree. While the expected time to inform such a vertex, given that all its neighbors are informed, is logarithmic, we need $\Omega(\log^2 n)$ rounds to do so with probability at least $n^{-1}$.
The following result shows that we need $\omega(\log n)$ rounds with probability $1-o(1)$ (see also~\tabref{SRG} for a survey).

\begin{table*}[bt]
    \footnotesize
    \begin{center}
    \scalebox{0.92}{
    \begin{tabular}{|c|c|}
    \hline
    \multicolumn{2}{|c|}{\bf Broadcast time}
    \\
    \multicolumn{1}{|c}{\bf Random model}
    & \multicolumn{1}{c|}{\bf Quasirandom model}
    \\
    \hline
    \hline
    $\Oh(\log^2 n)$ with probability $\geq 1-n^{-1}$ \cite{FPRU90}
        & \multirow{3}{*}{
\begin{tabular}{r}
 $\Oh(\log n)$ with probability $\geq 1-n^{-\gamma}$\ \ $\forall\gamma=\Oh(1)$ \\
         (\shortthmrefs{randomgraph}{main})
    \end{tabular}
}
    \\
        $\Omega(\log^2 n)$ with probability $\geq n^{-1}$ \cite{FPRU90}
        &
    \\
         $\Omega(\log(n) \,\log\log n)$ with probability $\geq 1-o(1)$ (\shortthmref{fullyrandom})
        &
    \\
    \hline
    \end{tabular}}
    \end{center}
    \caption{Summary of the broadcast times for
        almost all random graphs $G(n,p)$ with
        $p\,n=\log n+\omega(1)$ and $p\,n=\log n+\Oh(\log\log n)$.
    }
    \label{tab:SRG}
\end{table*}

\begin{thm}\label{thm:fullyrandom}
Let $p=(\log n + f(n))/n$, where $f(n)=\omega(1)$ and $f(n)\leq C \log \log n$ for some constant $C \geq 1$.
Then for almost all random graphs $G(n,p)$, the broadcast time of the fully random model is
   $\Omega(\log (n) \, \log \log n)$ with probability $1-o(1)$.
\end{thm}
\begin{proof}
 Fix an arbitrary vertex $v$. Then for any $x \geq 1$ we have,
 \begin{align*}
   \Pr{ \deg(v) \leq x} &\geq \Pr{ \deg(v) = x} \\ &= \binom{n-1}{x} \, p^x \, (1-p)^{n-1-x} \\
   		     &\geq \left( \frac{n-1}{x} \right)^x \left( \frac{\log n}{n}  \right)^x \, \left(1 - \frac{\log n+C \log \log n}{n} \right)^{n-1}.
		     \end{align*}
		     Now, using the fact that $\big(1-\frac{1}{n}\big)^{n-1} \geq e^{-1}$
		     twice gives
		     \begin{align*}
	 \Pr{ \deg(v) \leq x} &\geq  \left( \frac{n-1}{n} \right)^{x} \, \left(  \frac{\log n }{x}  \right)^x \, e^{-\log n - C \log \log n} \\
	 &\geq e^{-1} \, \left(  \frac{\log n}{x}  \right)^x \, e^{-\log n - C \log \log n}.
 \end{align*}

We now argue that with high probability, we have sufficiently many vertices of this small degree. The basic idea is to inspect the degree of the vertices in a careful manner. First, in order to verify whether a vertex $v_1$ has degree larger than $x$ or not, we only have to expose at most $x+1$ edges incident to $v_1$. Then, the next vertex we pick will be a vertex for which we have not exposed any edge so far. Using this way of exposing the vertices allows us to use a Chernoff bound and conclude that there are enough vertices of small degree.

More precisely, start with an arbitrary vertex $v_1 \in V$. In the first iteration, we check sequentially for all vertices $u \in V$ whether $\{v_1,u\} \in E$ until we know whether $\deg(v_1) \leq x$ holds or not. While we may have to check for up to $n-1$ vertices $u$ whether $\{v_1,u\}$ exists, we will never expose more than $x+1$ edges. This holds because after we have found $x+1$ edges incident to $v_1$, the event $\deg(v_1) \leq x$ does not hold. Then in the second iteration, we pick a new vertex $v_2 \neq v_1$ for which we have not exposed the existence of any edge (but we may already know that $\{v_2,v_1\} \notin E$). Again, we sequentially check for all vertices $u \in V$ whether $\{v_2,u\} \in E$ holds until we know whether $\deg(v_2) \leq x$ holds or not. Observe that we can continue in this manner as long as there is a new vertex $v_{i}$ for which we have not exposed the existence of any edge. Since in each iteration at most $x+1$ edges are exposed,
the number of vertices with no exposed edge is reduced by at most $x+2$ per iteration. As a consequence, the whole procedure can be run for at least $n/(x+2)$ iterations. In each iteration $1 \leq i \leq n/(x+2)$, we have
\begin{align*}
  \Pro{\deg(v_i) \leq x} &\geq e^{-1} \, \left(  \frac{\log n}{x}  \right)^x \, e^{-\log n - C \log \log n},
\end{align*}
by the same reasoning as above.

Let $X$ be the number of vertices with degree at most $x$. By the arguments above, it follows that $X$ is stochastically larger (cf. \defref{dominance} for a definition of stochastically larger) than
the sum of $n/(x+2)$ independent Bernoulli-random variables each of which has success probability $e^{-1} \, \left(  \frac{\log n}{x}  \right)^x \, e^{-\log n - C \log \log n}$.
Therefore, it follows by a Chernoff bound (\lemref{chernoff}) that
\begin{align}
 \Pr{ X \leq \tfrac{1}{2} \Ex{X} } &\leq
  e^{-  (1/2)^2 \Ex{X} /2}. \label{eq:chernoffapplied}
\end{align}

Now choose $x:=(\log n)^{\ee}$ for an arbitrary constant $0 < \ee < 1$. By the above, we obtain
\begin{align*}
  \Ex{X} &\geq \frac{n}{(\log n)^{\ee}+2} \, e^{-1} \left( \frac{\log n}{(\log n)^{\epsilon}} \right)^{(\log n)^{\epsilon}} \, e^{-\log n - C \log \log n} \\
 &\geq \tfrac{1}{3} (\log n)^{-\epsilon -C + (1-\epsilon) (\log n)^{\epsilon}  } = \left( \log n \right)^{\Omega( (\log n) ^{\epsilon})}.
\end{align*}
Plugging this into \eq{chernoffapplied}, we obtain
\begin{align*}
 \Pr{ X \leq (\log n)^{\Omega( (\log n)^{\epsilon} )} } &= o(1).
\end{align*}
By \cite[Lemma~1, Property 2]{CF07} we know that for almost all random graphs,
any two vertices with a degree of less than $\log n/20$ have a distance of at
least $\log n/(\log \log n)^2$ from each other. Hence, all neighbors of vertices
in $X$ have a degree of more than $\log n/20$. In particular, the time until a
vertex $u \in X$ gets contacted by a fixed neighbor $v \in N(u)$ is
stochastically larger
than a geometric random variable with parameter $\log n/20$. Hence the time
until $u$ gets contacted by any of its neighbors is stochastically larger than
the minimum of $\deg(u) \leq x = (\log n)^{\epsilon}$ independent such geometric
variables. Since any two vertices in $X$ have a distance of at least three,
these times are independent for all $u \in X$.

Now recall that $\RM(G)$ is the random variable describing the runtime of the fully random model. Further, let $\mathsf{Geo}(p)$ be the geometric distribution defined by $\Pro{\mathsf{Geo}(p) = i} = p \cdot (1-p)^{i}$ for any integer $i \geq 0$. Denoting with $\succeq$ ``stochastically larger" and using \lemref{mingeo}, we obtain
\begin{align*}
  \RM(G) &\succeq \max_{u \in X} \min_{v \in N(x)} \left \{ \mathsf{Geo}(20/\log n) \right \} \\
  &\succeq \max_{u \in X} \left\{ \mathsf{Geo}\left(1 - \prod_{v \in N(x)}(1 - 20/\log n) \right) \right\} \\
  &\succeq \max_{u \in X} \left\{ \mathsf{Geo}\left(1 - (1 - 20/\log n)^{(\log n)^{\epsilon}} \right) \right \} \\
  &\succeq \max_{i=1}^{\left( \log n \right)^{\Omega( (\log n) ^{\epsilon})}} \left\{ \mathsf{Geo}\left(1 - e^{-20 (\log n)^{\epsilon - 1}} \right) \right \}. \end{align*}
Hence\begin{align*}
 \Pro{ \RM(G) \leq t}
 &\leq \Pro{ \mathsf{Geo}\left(1 - e^{-20 (\log n)^{\epsilon - 1}} \right) \leq t } ^{ (\log n)^{\Omega( (\log n) ^{\epsilon}) } } \\
 &= \left( 1 - \left( e^{-20 (\log n)^{\epsilon - 1}}
 \right)^{t} \right) ^{ (\log n)^{\Omega( (\log n) ^{\epsilon}) } } \\
 &\leq \exp \left(- e^{-20 (\log n)^{\epsilon-1} t } \, (\log n)^{\Omega( (\log n) ^{\epsilon}) }
 \right).
\end{align*}
Setting $t = c \log n \log \log n $ with a sufficiently small  constant $c$ finally gives

\begin{align*}
 \Pro{ \RM(G) \leq t}
 &\leq  \exp \left(- (\log n)^{-20 c (\log n)^{\epsilon} } \,
        (\log n)^{\Omega( (\log n) ^{\epsilon})  }  \right) \\
        &= \exp \left( -(\log n)^{\Omega( (\log n) ^{\epsilon})  } \right).
 \qedhere
 \end{align*}
\end{proof}

\section{Quasirandom Rumor Spreading on Hypercubes}
\label{sec:hypercube}

In this section we analyze the quasirandom model on hypercubes. We prove that the quasirandom model informs all vertices in $\Oh(\log n)$ rounds with high probability.
This extends a corresponding runtime bound of $\Oh(\log n)$ for the fully random model in~\cite{FPRU90}. The difficulty in our analysis is that the hypercube is not an expanding graph (cf.~\thmref{HCnotexpanding}), and also an application of the bound of \thmref{degdiambound} yields only a much weaker upper bound of $\Oh(\log^2 n)$.

We now state and prove our runtime bound for the quasirandom model on hypercubes. Finally, we will also examine the failure probability more closely to reveal that there is again a slight superiority of the quasirandom model over the fully random model (\secref{failure}).

\newcommand{\stageone}{3d} \newcommand{\stagetwo}{6d} \newcommand{\stagethree}{1033d} 
\begin{thm}
    \label{thm:hypercube}
    The broadcast time of the quasirandom model on the hypercube is
    $\Oh(\log n)$ with probability $1-n^{-\Omega(\log n)}$.
\end{thm}

Similarly to the proof for expanding graphs in \secref{analysis},
the analysis consists of a forward part and backward
part. While the analysis of the forward part borrows several concepts from the analysis of the fully random model \cite{FPRU90}, the idea of analyzing the
process in reversed order was not used in \cite{FPRU90}.

\begin{figure}
    \center
    \scalebox{.8}{\input{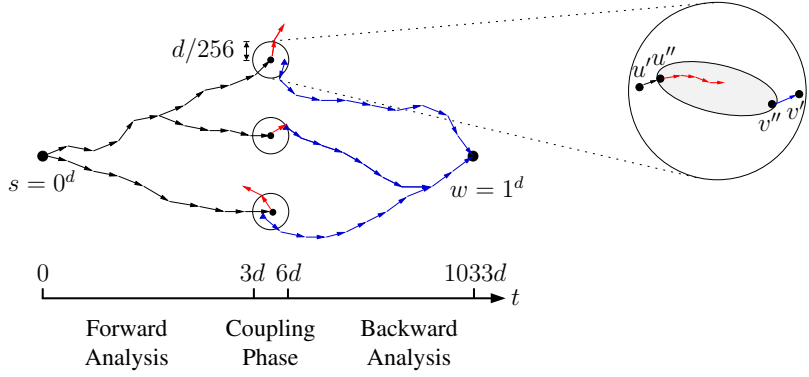}}
    \caption{The left side contains a sketch of the proof of \thmref{hypercube}.
        The black circles represent $I'_{\stageone}$, and the triangles represent
        $\bigcup_{v\in I'_{\stageone}}\Phi(v)$. The right side illustrates the analysis of the
        coupling part. We find two vertices $v''$ and $u''$ such that every
        shortest path between them is included in a subcube of vertices whose initially-contacted
        neighbors are not exposed.}
    \label{fig:sketchcube}
\end{figure}

The forward part informs sufficiently many vertices in $\Oh(\log n)$ time. The backward part shows that if there is an uninformed vertex, then $\Oh(\log n)$
steps earlier every ball of small radius in the hypercube contains at least
one uninformed vertex. To prove that one of these uninformed vertices gets
informed eventually, we need a third part in between, which we call coupling.
A graphical illustration of our proof can be found in \figref{sketchcube} on page~\pageref{fig:sketchcube}.

To formally prove \thmref{hypercube}, we assume that the following
three lemmas hold.  We state them here
and prove them in the remainder of this section.
Recall that $n=2^d$.

\begin{lem}
\label{lem:cubeforward}
    The probability that the quasirandom rumor spreading process started in a fixed vertex $s$
    informs $2^{d/6}$ vertices in $\stageone$ steps is at least
    $1 - n^{-\Omega(\log n)}$.
\end{lem}

Let $s=0^d$ be initially informed.
By \lemref{cubeforward}, at least
$2^{d/6}$ vertices get informed in $\stageone$
with probability at least $1-n^{-\Omega(\log n)}$.
Now fix an arbitrary vertex $w \in V$.
Recall that $U_{[t_1,t_2]}(w)$ is the set of vertices that reach
the vertex $w$ within the time interval $[t_1,t_2]$ (cf.~definition on page~\pageref{def:reach}).

\begin{lem}
\label{lem:cubebackward}
    For any vertex $w$ and $t_2=\stagethree$, with
    probability at least $1-n^{-\Omega(\log n)}$, there is for every vertex $v$ a vertex
    $u(v) \in U_{[\stagetwo,t_2]}(w)$ with $\dist(u,v) \leq d/256$.
\end{lem}

By applying \lemref{cubebackward}, there is with probability at least $1-n^{-\Omega(\log n)}$ for each $v \in I_{\stageone}$ a vertex $u(v) \in U_{[\stagetwo,t_2]}(w)$ with $\dist(u,v) \leq d/256$.

\begin{lem}
\label{lem:cubecoupling}
    Let $s$ be the initially informed vertex
    and $w$ be an arbitrary vertex.
    Assume that the following two conditions hold:
    \begin{itemize}
    \setlength{\itemsep}{0pt}
    \setlength{\parskip}{0pt}
    \item
        there are at least $2^{d/6}$ informed vertices at step $\stageone$ and
    \item
        there is for every vertex $v$ a vertex $u(v) \in
        U_{[\stagetwo,t_2]}(w)$ with $\dist(u,v) \leq d/256$
        and $t_2=\stagethree$.
    \end{itemize}
    Then with probability $1 - e^{-\poly(n)}$,
    at least one vertex in $U_{[\stagetwo,t_2]}(w)$
    is informed at step~$\stagetwo$.
\end{lem}

Now if the two former conditions hold, \lemref{cubecoupling} implies that a vertex in $U_{[\stagetwo,t_2]}(w)$ gets informed with (conditional) probability at least $1-n^{-\Omega(\log n)}$.
By definition this implies that the vertex $w$ gets informed at step $t_2$.
Taking the union bound over the success of the forward and backward part (\lemref{cubeforward} and \lemref{cubebackward}),
it follows that at step $t_2$ the vertex $w$ gets informed with probability at least $1 - n^{-\Omega(\log n)}$.
Taking the union bound over all possible vertices $w \in V$ yields \thmref{hypercube}.


\subsection{Proof of the Forward Analysis}
\label{sec:cubeforward}

In this section we prove \lemref{cubeforward}.

\begin{proof}[Proof of \lemref{cubeforward}]
    By symmetry we may assume that $s=0^d$ is initially informed.
    Let $L_i$ be the set of vertices with $\|x\|_1=i$.
    Note that after two phases of $d$ steps each,
    we have $I_{2d} = \{s \} \cup L_1 \cup L_2$.

    Consider some time-step $t \geq 2d$. Assume that all initially-contacted neighbors of~$I_{t} \cap L_i$ are still to be chosen \uar\
    for $i\geq 2$.
    Notice that the number of edges between $I_{t} \cap L_{i}$ and $L_{i+1}$ is
    $ | E(I_{t} \cap L_{i},L_{i+1})| = \sum_{v \in L_{i+1}} \deg_{I_{t} \cap L_{i}}(v) = |I_{t} \cap L_i| \, (d-i)$.
    Our goal is to show that a large set of vertices in $L_{i+1}$ will be informed after a phase of $4$ additional steps.
    The probability that a vertex~$v \in L_{i+1}$ is still uninformed after this phase is
    \begin{align*}
      \Pr{v \not\in I_{t+4}}
      &\leq \prod_{u \in \Gamma(v) \cap I_{t} \cap L_{i}} \left(1 - \frac{4}{d} \right)
      = \left(1 - \frac{4}{d} \right)^{\deg_{I_{t} \cap L_i}(v)}.
    \end{align*}
    By linearity of expectations we get
   \begin{align*}
        \Ex{ | I_{t+4} \cap L_{i+1}| }
        &= \sum_{v \in L_{i+1}} \Pr{v \in I_{t+4}}
        \geq \sum_{v \in L_{i+1}} 1 - \left(1 - \frac{4}{d} \right)^{\deg_{I_t \cap L_i}(v)} \\
        &\geq \sum_{v \in L_{i+1}} 1 - \exp\bigg(-\frac{ 4 \,\deg_{I_t \cap L_{i}}(v)}{d}\bigg).
   \end{align*}
   Let us now assume that $1 \leq i \leq d/4-1$.
   Then since $\deg_{I_t \cap L_i}(v) \leq i+1$ for $v \in L_{i+1}$ and $1+\frac{x}{2} \geq e^x$ for any $-1 \leq x \leq 0$, we get
   \begin{align*}
        \Ex{ | I_{t+4} \cap L_{i+1}| }
        &\geq \sum_{v \in L_{i+1}} \frac{2 \deg_{I_t \cap L_i}(v)}{d}
        = \frac{2}{d} \, |I_{t} \cap L_i|\, (d-i)
        = 2 \, \frac{d-i}{d} \, |I_{t} \cap L_i|.
    \end{align*}
   Since any vertex of $|I_t \cap L_i|$ can only inform at most $4$ vertices within $4$ steps, an application of Azuma's inequality (cf.~\lemref{Azuma}) gives, for any constant $0 < \epsilon \leq 2/3$,
    \begin{align*}
       &\Pr{ |I_{t+4} \cap L_{i+1}| \leq (2-\epsilon) \,\frac{d-i}{d} \, |I_{t} \cap L_i| }\\
       &\qquad\leq \exp \biggl(- \frac{  (\epsilon \, \frac{d-i}{d} \, |I_{t} \cap L_i|)^2}{16 \,|I_{t} \cap L_i|} \biggr)
       = \exp(-\Omega(d^2))
       = n^{-\Omega(\log n)},
     \end{align*}
     as long as $|I_t \cap L_i| \geq \frac{d\,(d-1)}{2}$
      holds. Observe that if the condition $|I_t \cap L_i| \geq \frac{d\,(d-1)}{2}$ holds initially, then
         $
         |I_{t+4} \cap L_{i+1}| \geq (2-\epsilon) \frac{d-i}{d} |I_{t} \cap L_i|$ implies that $|I_{t+4} \cap L_{i+1}| \geq \frac{d\,(d-1)}{2}$, since $(2-\epsilon) \frac{d-i}{d} \geq (2-\epsilon) \frac{3}{4} \geq 1$ by definition of $i$ and $\epsilon$.

     Recall that we first spent $2d$ steps in the first two phases to inform $L_{2}$ completely. Then in the analysis above, we spent, for each level $i$ with $2 \leq i \leq d/4 - 1$, a phase of exactly $4$ steps. Hence the total time consumption is
     \[
       2 d + (d/4 - 2) \cdot 4 \leq 3d.
     \]
     Now taking the union bound over all levels $2 \leq i \leq d/4 - 1$,
     with probability $1-(d/4-1) \, n^{-\Omega(\log n)} = 1 - n^{-\Omega(\log n)}$ it holds that
     \begin{align*}
       | I_{4d} \cap L_{d/4} | &\geq \frac{d\,(d-1)}{2} \, \prod_{i=2}^{d/4-1} \left( (2-\epsilon) \,\frac{d-i}{d} \right) \\
       &= \frac{d\,(d-1)}{2} \, \left(2 - \epsilon \right)^{d/4-2} \, \prod_{i=2}^{d/4-1} \left( 1 - \frac{i}{d} \right).
       \end{align*}
       We now use the fact that $(1-x)^{1/x}$ is non-increasing in $0 < x < 1$, implying $(1-x) \geq 4^{-x}$ for any $x \leq 1/4$. Plugging this into the previous inequality yields
       \begin{align*}
    | I_{4d} \cap L_{d/4} |   &\geq  \left(2 - \epsilon \right)^{d/4} \, 4^{-\sum_{i=2}^{d/4-1} \frac{i}{d}  }
       \geq \left(2 - \epsilon \right)^{d/4} \, 4^{-d/32 }
       \geq 2^{d/6},
     \end{align*}
     if $\epsilon > 0$ is a suf{}ficiently small constant.
\end{proof}


\subsection{Proof of the Backward Analysis}
\label{sec:cubebackward}

In this section we prove \lemref{cubebackward}. We shall use the notation that $x[j]$ denotes the $j$-th bit of a vertex $x \in V$.

\begin{proof}[Proof of \lemref{cubebackward}]
    We will now analyze the propagation of the rumor in the
    reverse order. Due to the symmetry of $H$, we may restrict our attention to the case~$w=1^d$.

    Let us first consider the case where $v=0^d$.
    So we have to show that $U_{[\stagetwo,t_2]}(w)$ contains a vertex~$u$
    such that $\dist(0^d,u) \leq d/256$ with probability at least $1-n^{-\Omega(\log n)}$. 
    In order to achieve such a large success probability, we will construct $d/512$ vertex-disjoint paths that start from a vertex in $\Gamma(w)$ and move towards the vertex $v$. For each neighbor of $w$ which differs from $w$ in one of the last $d/512$ bits, we associate a path starting from that vertex and moving towards the vertex $v$. The disjointness is ensured by not allowing the path to change any of the last $d/512$ bits.

   First note that $U_{[t_2-d,t_2]}(w) \supseteq \Gamma(w)$, since within a time interval of $d$ steps, every neighbor of $w$ contacts $w$.
    Let $\mathcal{J} := [(511/512)\,d,d]$. For each $j \in \mathcal{J}$, we define a set of vertices
%
\begin{align*}
 V(j) &:=
 \left\{x\in \{0,1\}^d
 \text{\ with $x[j]=0$ and $x[i]=1$ for $i \in [(511/512) \,d,d] \backslash \{j\}$ }
 \right\}.
 \end{align*}

    For each $j \in \mathcal{J}$ we consider a path $P(j)=(v_1,v_2,\ldots,v_\ell) \subseteq V(j)$
    of length $\ell:=(255/256)\,d$
    which is defined inductively as follows:
    \begin{itemize}
      \setlength{\itemsep}{0pt}
      \setlength{\parskip}{0pt}
      \item The first vertex of $P(j)$ is defined by $v_1 \in \Gamma(w) \cap V(j)$.
      \item If $s_i$ denotes the time-step when $P(j)$ has reached the vertex $v_i$, then $P(j)$ is extended to a vertex $v_{i+1} \in \Gamma(v_i) \cap V(j)$ with $\|v_{i+1}\|_1=d-i-1$ such that $v_{i+1}$ is the last vertex before time-step $s_i$ that contacts $v_{i}$.
    \end{itemize}

    \noindent Fix an arbitrary $j \in \mathcal{J}$ and consider the path $P(j)$. Recall that $v_1 \in U_{[t_2-d,t_2]}(w)$.
    Fix any $i$ with $1 \leq i \leq \ell$ and consider the vertex $v_{i}$. Note that there are $d-i-(1/512)\,d$ vertices $u \in \Gamma(v_i) \cap V(j)$ with $\|u\|_1 = \|v_i\|_1 - 1$.
    Let us denote by $\Delta_i(v_{i})$ the waiting time (going back in time) until such a fixed vertex $u$ contacts $v_{i}$, in symbols,
    \[
      \Delta_i(u,v_i) := s_i - \max \{ s \leq s_i - 1 \colon u \in U_{[s,s_i]}(v_i) \}.
    \]

    Note that $\Delta_i(u,v_i)$ is a uniform random variable in $\{1,\ldots,d\}$. In particular, the distribution is the same for every $u$ and since the initially-contacted neighbors are chosen independently and uniformly at random, $\left\{ \Delta_i(u,v_i) \colon u \in \Gamma(v_i) \cap V(j), \|u\|_1 = \|v_i\|_1 - 1 \right\}$ is a set of mutually independent random variables.
    The waiting time $\Delta_i$ until the first vertex $u \in \Gamma(v_i) \cap V(j)$ with $\|u\|_1 = \|v_i\|_1 - 1$ contacts $v_{i}$ satisfies
     \begin{align*}
     \Delta_i &:= \min_{\substack{u \in \Gamma(v_i) \cap V(j) \colon \\ \|u\|_1 = \|v_i\|_1 - 1}} \Delta_i(u,v_i).
    \end{align*}

    To bound this random variable, let $X_{i,u}\sim\mathsf{Geo}(1/d)$, that is,
    a geometric random variable with parameter $1/d$.
    By \lemref{mingeo}, the minimum of $d-i-(1/512)\,d$ independent geometric
    random variables with parameter $1/d$ is itself a geometric
    random variable~$X_{i}$ with parameter
    \[
    1- \left(1- \frac{1}{d} \right)^{d-i-(1/512)\,d} \geq 1 - \exp \left( -1/512 \right) \geq 1 - \frac{1}{1/512 + 1} = \frac{1}{513}. \]
    Hence with \textquotedblleft{}$\preceq$\textquotedblright{} denoting \textquotedblleft{}stochastically smaller than\textquotedblright{} we obtain by \lemref{totalbanal} that
    \begin{align*}
        \Delta_i &= \min_{\substack{u \in \Gamma(v_i) \cap V(j) \colon \\ \|u\|_1 = \|v_i\|_1 - 1}}
            \Delta_i(u,v_i)
        \preceq \min_{\substack{u \in \Gamma(v_i) \cap V(j) \colon \\ \|u\|_1 = \|v_i\|_1 - 1}}
            X_{i,u}
        = X_i.
     \end{align*}
     Hence the time $\Delta(j):=\sum_{i=1}^{\ell} \Delta_i$ until we reach the
     end of $P(j)$ is stochastically smaller than $\sum_{i=1}^{\ell} X_i$, where the $X_i$'s are independent geometric
     random variables with parameter $1/513$.

     Let us first note that $\Ex{X_i} \leq 513$ and therefore with $X:=\sum_{i=1}^{\ell} X_i $,
     \[
        \Ex{X} = \sum_{i=1}^{\ell} \Ex{X_i} \leq 513\, d.
    \]
     Now we apply a Chernoff bound for a sum of independent geometric random
     variables (\lemref{chernoffgeo} with $\epsilon:=1$) to obtain
     \begin{align*}
        \Pr{ X \geq 1026\, d  } &\leq \exp \left(- \frac{1}{4} \ell \right),
     \end{align*}
     and since $\Delta(j) \preceq X$,
     \begin{align*}
        \Pr{ \Delta(j) \geq 1026 \,d }
        &\leq   \exp \left(- \frac{1}{4} \ell \right).
     \end{align*}

    Hence with probability $1- \exp (- \frac{1}{4} \ell )$, the endpoint of a path $P(j)$ for a fixed $j$
    contacts~$w$ within the time interval $[\stagetwo,t_2]$.

    Note that $\{\Delta(j) \colon j \in \mathcal{J}  \}$ is a set
    of independent random variables, since
    for any $j_1,j_2\in\mathcal{J}$ with $j_1\neq j_2$,
    the vertex sets $V(j_1)$ and $V(j_2)$ are disjoint. Using this
    independence, we can lower bound the probability that there is a vertex $u$
    with $\|u\|_1 \leq d/256$ and $u \in U_{[\stagetwo,t_2]}(w)$ by
    \begin{align*}
        1 - \left( \exp \left(- \frac{1}{4} \ell \right) \right)^{|\mathcal{J}|}
        &\geq 1 - e^{-\Omega(d^2)}
        = 1 - n^{-\Omega(\log n)}.
    \end{align*}
    So far, we have considered the case where $v=0^d$.
    With the same arguments, we can prove that for an arbitrary vertex $v$
    there is a vertex $u(v)$ satisfying $\dist(u(v),v) \leq d/256$
    and $u(v) \in U_{[\stagetwo,t_2]}(w)$ with probability $1-n^{-\Omega(\log n)}$.
    It follows by a union bound that with probability $1-n^{-\Omega(\log n)}$, there is for every vertex $v \in V(G)$
    a vertex $u(v) \in U_{[\stagetwo,t_2]}(w)$ with
    $\dist(v,u(v)) \leq d/256$.
\end{proof}


\subsection{Proof of the Coupling Part}
\label{sec:cubecoupling}

In this section we prove \lemref{cubecoupling}.

\begin{proof}[Proof of \lemref{cubecoupling}]
    Let $w$ be an arbitrary, fixed vertex.
    By the first condition in \lemref{cubecoupling}, we have $|I_{\stageone}| \geq 2^{d/6}$. By definition of the hypercube, there are for every vertex $u$ exactly $\sum_{k=0}^{d/64} \binom dk$ vertices with distance at most $d/64$ to $u$. Hence there is subset $I'_{\stageone} \subseteq I_{\stageone}$ such that two vertices in $I'_{\stageone}$ have distance at least $d/64$ from each other which is of size
    \[
        \frac{2^{d/6}}{\sum_{k=0}^{d/64} \binom dk}
        \geq \frac{2^{d/6}}{(64e)^{d/64}}
        \geq \frac{2^{d/6}}{ \left(2^{8} \right)^{d/64} }
        = 2^{d/24},
    \]
    where we have used the inequality $\sum_{i=0}^{m} \binom{n}{i} \leq \big(\frac{e\,n}{m}\big)^m$.

    By our second condition in \lemref{cubecoupling}, there is for each vertex $v \in I'_{\stageone}$ at least one vertex $u=u(v) \in U_{[\stagetwo,t_2]}(w)$ such that $\dist(u,v) \leq d/256$.

    Let $\Phi\colon I'_{\stageone} \rightarrow U_{[\stagetwo,t_2]}(w)$
    be a function that assigns each vertex $v \in I'_{\stageone}$ a vertex $u=u(v) \in U_{[\stagetwo,t_2]}(w)$ such that $\dist(u,v) \leq d/256$. Using the fact that two vertices in $I'_{\stageone}$ have distance at least $d/64$ from each other, we observe that $\Phi$ is an injective function.

    Let us now fix a pair of vertices $v \in I'_{\stageone}$ and $\Phi(v) \in U_{[\stagetwo,t_2]}(w)$. Note that the set of all shortest paths between $v$ and $\Phi(v)$ form a subcube $H'=H'(v,\Phi(v))$ whose dimension is equal to the distance between $v$ and $\Phi(v)$. Now choose a pair of vertices $v' \in H' \cap I_{\stageone}$ and $u' \in H' \cap U_{[\stagetwo,t_2]}(w)$ such that $\dist(v',u')$ is minimized. Our aim is to lower bound the probability that $v'$ reaches $u'$ within the time interval $[\stageone,\stagetwo]$.

    First let us assume that $\dist(v',u') \leq 3$. In this case, $u'$ is informed within $3d$ steps with probability $1$. Otherwise, we have $\dist(v',u') \geq 4$. In this case, let $v'' \in \Gamma(v')$ and $u'' \in \Gamma(u')$ be two vertices such that $\dist(v'',u'') = \dist(v',u') - 2$. Note that $v'' \in I_{4d}$ and $u'' \in U_{[5d,t_2]}(w)$. By our construction, every vertex on a shortest path between $v''$ and $u''$ (except $u''$) has distance at least one to $I_{\stageone}$ and distance at least two to $U_{[\stagetwo,t_2]}(w)$. Hence for each vertex on such a shortest path, the initially-contacted neighbor is still chosen uniformly at random and independently of all other vertices.

    Similarly to the proof of \lemref{cubebackward}, we lower bound the probability that there exists a path $P=P(v)=(v_1 =v'',v_2,\ldots,v_{\dist(v'',u'')}= u'')$ which satisfies the following two conditions for any $1 \leq i < \dist(v'',u'')$:
    \begin{itemize}
        \setlength{\itemsep}{0pt}
        \setlength{\parskip}{0pt}
     \item $v_{i+1}$ is closer to $u''$ than $v_{i}$ and
     \item $v_{i}$ informs $v_{i+1}$ at step $4d + i$.
    \end{itemize}

    \noindent
    Note that once the rumor has reached a vertex $v_{i}$ for the first time, the vertex $v_{i}$ forwards it to a vertex $v_{i+1}$ closer to $v''$ with probability at least $(\dist(v'',u'')-i+1)/d$.
    Repeating this argument gives the following lower bound for the existence of $P$:
    \[
        \prod_{i=1}^{d/256} \frac{i}{d}
        = \frac{(d/256)!}{d^{d/256}}
        \geq \frac{d^{d/256}}{(768\,d)^{d/256}}
        \geq \left( 2^{-10} \right)^{d/256}
        \geq 2^{-d/25},
    \]
    where we have used the fact that $n! \geq (n/3)^n$ for any integer $n$ in the left inequality.

    Our next claim is that $\{ \mbox{$P(v)$ exists} \colon v \in I'_{\stageone} \}$ is a set of mutually independent events. In order to prove this, let us consider two arbitrary vertices $v_1, v_2 \in I'_{\stageone}$, $v_1 \neq v_2$. Recall that by definition of $I'_{\stageone}$, $\dist(v_1,v_2) \geq d/64$. Since every vertex on a shortest path between $v_i$ and $\Phi(v_i)$ has a distance of at most $d/256$ to $v_i$, it holds by the triangle inequality that the two paths $P(v_1)$ and $P(v_2)$  always have a  distance of at least $d/128$ from each other, which proves the claimed independence.

    Using this, we can lower bound the probability that at least one $P(v)$ exists by
    \[
        1 - \Bigl(1- 2^{-d/25} \Bigr) ^ { 2^{d/24}} \geq 1 - \exp \left(-2^{d/(24 \cdot 25)} \right) = 1 - \exp \left( -\poly(n)  \right).
    \]
    If there is a $v \in I'_{\stageone}$ for which $P(v)$ exists, then we know that there is a vertex $v''(v) \in I'_{4d}$ which reaches a vertex $u''(v) \in U_{[5d,t_2]}(w)$ within the time interval $[\stageone,\stagetwo]$. This implies that $u(v) \in U_{[\stagetwo,t_2]}(w)$ is informed at step $\stagetwo$, and as a consequence, $w$ will become informed at step $t_2$.
\end{proof}

\subsection{Failure Probability}\label{sec:failure}

We now examine the probabilities in the runtime bounds for the hypercube more closely. Recall that the runtime bound of $\Oh(\log n)$ for the quasirandom model holds with probability at least $1-n^{-\Omega(\log n)}$. In the fully random model, however, a fixed vertex remains uninformed for $x$ steps with probability at least
$(1-1/d)^{d x} \geq 4^{-x}$. Hence the runtime of the fully random model is at least $\rho \cdot \log_2 n$ with probability at least $n^{-2 \rho}$ for any value of $\rho \geq 1$. Hence if $\rho = (c/2) \log_2 n$ for some constant $c > 0$, this shows that the time for the fully random model to inform all $n$ vertices with probability at least $1-n^{-c \log_2 n}$ is at least $(c/2) (\log_2 n)^2=\Omega(\log(n)^2)$. This should be compared with our upper bound of $\Oh(\log n)$ for the quasirandom model, which holds with probability at least $1-n^{-\Omega(\log n)}$.


\section{Conclusion and Outlook}\label{sec:conclusion}

In this paper, we proposed and investigated a quasirandom analogue of the
classical push model for spreading a rumor to all vertices of a
network.

We showed that for many network topologies, after $\Theta(\log n)$ iterations all vertices
are informed with probability $1 - \Oh(\poly(n))$. Hence the quasirandom model achieves
asymptotically the same bounds as the random one, or even better ones (\eg\ for
random graphs with $p$ close to $\log(n)/n$).

This work is also interesting from the methodological point of view.
Our proofs
show, in particular, that the difficulties usually invoked by highly dependent
random experiments can be overcome.
From the general perspective of using randomized methods in computer science,
our results, as a number of other recent results, can be viewed as suggesting
that choosing the right dose of randomness might be a fruitful topic for
further research.

An interesting open problem is to analyze the quasirandom push model on other graph classes. A natural candidate would be the class of regular graphs with constant conductance, for which it is known that the classical push model spreads a rumor in $\Oh(\log n)$ rounds \cite{Giakkoupis11,CLP10b}. Another interesting target are preferential attachment graphs. Here~\cite{DoerrFF11} have shown that the fully random push-pull model has a broadcast time of $\Theta(\log n)$, whereas the variant with contactees chosen uniformly at random from all neighbors except the previous contactee has a broadcast time of only $\Theta(\log n / \log\log n)$. Since the quasirandom protocol automatically avoids the previous contactee, it seems likely that it also has this superior broadcast time.

Note however that it is not true that the quasirandom model always performs at least as good as the fully random model. For instance, consider the graph consisting of two cliques of size $n/2-1$ and an extra vertex which is connected to all other $n/2-2$ vertices. On this graph the fully random model spreads a rumor in $\Oh(\log n)$ rounds with high probability, whereas the quasirandom model needs $\Omega(n)$ rounds with probability at least $1/4$ for appropriately chosen lists.

%


\bibliographystyle{abbrvnat}

\newcommand{\FOCS}[2]{#1 IEEE Symposium on Foundations of Computer Science (FOCS)}
\newcommand{\STOC}[2]{#1 ACM Symposium on Theory of Computing (STOC)}
\newcommand{\SODA}[2]{#1 ACM-SIAM Symposium on Discrete Algorithms (SODA)}
\newcommand{\ICALP}[2]{#1 International Colloquium on Automata, Languages, and Programming (ICALP)}
\newcommand{\PODC}[2]{#1 ACM-SIGOPT Principles of Distributed Computing (PODC)}
\newcommand{\STACS}[2]{#1 International Symposium on Theoretical Aspects of Computer Science (STACS)}
\newcommand{\SPAA}[2]{#1 ACM Symposium on Parallel Algorithms and Architectures (SPAA)}
\newcommand{\MFCS}[2]{#1 International Symposium on Mathematical Foundations of Computer Science (MFCS)}
\newcommand{\ISAAC}[2]{#1 International Symposium on Algorithms and Computation (ISAAC)}
\newcommand{\WG}[2]{#1 Workshop of Graph-Theoretic Concepts in Computer Science (WG)}
\newcommand{\SIROCCO}[2]{#1 International Colloquium on Structural Information and Communication Complexity (SIROCCO)}
\newcommand{\IPDPS}[2]{#1 International Parallel and Distributed Processing Symposium (IPDPS)}
\newcommand{\DISC}[2]{#1 International Symposium on Distributed Computing (DISC)}
\newcommand{\RANDOM}[2]{#1 International Workshop on Randomization and Computation (RANDOM)}
\newcommand{\ESA}[2]{#1 European Symposium on Algorithms (ESA)}
\newcommand{\IPPS}[2]{#1 IEEE International Parallel Processing Symposium (IPPS)}
\newcommand{\TAMC}[2]{#1 Annual Conference on Theory and Applications of Models of Computation (TAMC)}


\appendix

\newpage
\section{Probabilistic Tail Bounds Used for our Analysis}

As our analysis heavily relies on probabilistic tails bounds,
we summarize them here for reference.  The following bound
can be found, e.g., in the textbook of \citet{MU05}.

\medskip


\begin{lem}[Chernoff bounds for sums of Bernoulli variables]
    \label{lem:chernoff}
    Let $X_i, 1 \leq i \leq n$, be independent random variables.
    Let $X=\sum_{i=1}^n X_i$, $0<p<1$ and $\delta>0$.
    If $\Pr{X_i=1}=p$ and $\Pr{X_i=0}=1-p$ for all~$i \in \{1, \ldots, n\}$, then
    \begin{align*}
        \Pr{X \le (1-\delta) \Ex{X}} &\le \exp\left(-\delta^2 \Ex{X} / 2\right),\\
        \Pr{X \ge (1+\delta) \Ex{X}} &\le \exp\left(-\min\{\delta,\delta^2\} \Ex{X}/3\right).
    \end{align*}
\end{lem}

\noindent
We also use the following concentration bound, which is also called the method of bounded differences~\cite[Lemma~1.2]{mcdiarmid89}.
\begin{lem}[Azuma's inequality]
    Let $X_i\colon \Omega_i\to\R$, $1\leq i\leq n$, be mutually independent random variables.
    Let 
    $f\colon \prod_{i=1}^n \Omega_i \to\R$ satisfy
    the Lipschitz condition
    \begin{align*}
        | f(\x) - f(\x') | \,&\leq\, c_i\\
    \intertext{where $\x$ and $\x'$ differ only in the $i$-th coordinate, $1\leq i\leq n$.
        Let $Y$ be the random variable $f(X_1,\ldots,X_n)$.
        Then for any $t\geq0$,}
        \Pr{ Y > \Ex{Y} + t} \,&\leq\, \exp(-2t^2/\tsum_{i=1}^n c_i^2).
    \end{align*}
    \label{lem:Azuma}
\end{lem}

%

We shall also use the concept of stochastic domination between random variables.
\begin{definition}\label{def:dominance}
A random variable $X$ is \emph{stochastically smaller} than $Y$, if for all $k \in \R$, $\Pr{ X \geq k} \leq \Pr{ Y \geq k}$.
In this case, we also write $X \preceq Y$.
\end{definition}

\medskip
We list two obvious facts about stochastic domination.
\begin{lem}\label{lem:totalbanal}
Let $X_1, X_2$ be two independent random variables and let $Y_1,Y_2$ be two additional independent random variables with $X_1 \preceq Y_1$ and $X_2 \preceq Y_2$. Then,
\begin{itemize}
    \setlength{\itemsep}{0pt}
    \setlength{\parskip}{0pt}
\item $X_1 + X_2 \preceq Y_1 + Y_2$ and
\item $\min\{X_1,X_2\} \preceq \min \{Y_1,Y_2\}$.
\end{itemize}
\end{lem}

\medskip

We continue with a simple fact about the geometric distribution.
\begin{lem}\label{lem:mingeo}
Let $X_1,X_2,\ldots,X_n$ be $n$ independent geometric random variables each with parameter $0 < p < 1$. Then $X:= \min_{i=1}^n X_i$ is a geometric random variable with parameter $(1 - (1-p)^{n})$.
\end{lem}
%

\medskip

We use the following standard Chernoff bound for sums of geometric random variables from~\cite[Problem~3.6]{DP09}.
\begin{lem}[{Chernoff bound for sums of geometric variables
}]
    \label{lem:chernoffgeo}
    Let $Y_1, Y_2, \ldots, Y_n$ be independent geometric random variables,
    each with parameter $p > 0$. Let $Y:=\sum_{i=1}^n Y_i$. Then for any $\epsilon > 0$,
\begin{align*}
 \Pr{ Y \geq (1+\epsilon) \,\frac{n}{p}} &\leq \exp\left(-\frac{\epsilon^2}{2\,(1+\epsilon)} \, n \right).
\end{align*}
\end{lem}




\end{document}